\documentclass[10pt,a4paper]{article}
\usepackage[utf8]{inputenc}
\usepackage[left=1.8cm,right=1.8cm,top=2cm,bottom=2cm]{geometry}

\usepackage{makecell}
\usepackage{forest}
\usepackage{setspace}
\usepackage{textcomp}

\usepackage{hyperref}
\usepackage{url}

\usepackage[electronic]{ifsym}
\usepackage{indentfirst,latexsym}
\usepackage{bm}
\usepackage{amsmath,amsfonts,amsthm,amscd,amssymb}
\usepackage{pifont}
\usepackage{url}

\usepackage{float}

\usepackage{longtable}

\usepackage{graphicx}
\usepackage{subfigure}

\usepackage[all,pdf]{xy}

\usepackage{booktabs}
\usepackage{multirow}
\usepackage{multicol}
\usepackage{stfloats}  
\usepackage{orcidlink}

\DeclareMathOperator{\dif}{d}  
\DeclareMathOperator{\dist}{\mathrm{dist}}

\renewcommand{\vec}[1]{\mathbf{#1}}

\newcommand{\Real}{\mathbb{R}}

\newcommand{\secode}[2]{\frac{\dif^2 {#1}}{\dif {#2}^2}}
\newcommand{\set}[1]{\left\{ #1 \right\}}

\newcommand{\abs}[1]{\left| #1 \right|}
\newcommand{\braket}[2]{ \langle #1 | #2 \rangle}
\newcommand{\norm}[1]{\left\lVert #1 \right\rVert}

\newcommand{\trsp}[1]{{#1}^\textsf{T}}

\newcommand{\ES}[3]{\mathbb{#1}^{{#2}\times {#3}}}     

\DeclareMathOperator{\AGM}{AGM}

\newcommand{\physdim}[1]{[\texttt{#1}]}

\newcommand{\scru}[2]{{#1}^{\mathrm{#2}}}
\newcommand{\scrd}[2]{{#1}_{\mathrm{#2}}}

\usepackage{algorithm}         
\usepackage{algorithmicx}
\usepackage{algpseudocode}
\usepackage{float}

\newcommand{\Fig}{\textbf{Figure}~}
\newcommand{\Tab}{\textbf{Table}~}

\algnewcommand\algorithmicswitch{\textbf{switch}}
\algnewcommand\algorithmiccase{\textbf{case}}
\algnewcommand\algorithmicdefault{\textbf{default}}

\algdef{SE}[SWITCH]{Switch}{EndSwitch}[1]{\algorithmicswitch\ #1\ \algorithmicdo}{\algorithmicend\ \algorithmicswitch}%
\algdef{SE}[CASE]{Case}{EndCase}[1]{\algorithmiccase\ #1}{\algorithmicend\ \algorithmiccase}%
\algdef{SE}[DEFAULT]{Default}{EndDefault}[1]{\algorithmicdefault\ #1}{\algorithmicend\ \algorithmicdefault}%

\makeatletter
\renewcommand{\ALG@name}{Algorithm}

\usepackage{caption}
\usepackage{listings}
\newcommand{\cpfile}[1]{\texttt{#1}}

\newcommand{\PtrToFun}[1]{\texttt{#1}}
\newcommand{\ProcName}[1]{\textsc{#1}}

\newtheorem{thm}{Theorem}
\newtheorem{StemProb}[thm]{STEM Problem}

\author{Hong-Yan Zhang\orcidlink{0000-0002-4400-9133}\thanks{Corresponding author, e-mail: hongyan@hainnu.edu.cn}, Yu Zhou, Yu-Tao Li, Fu-Yun Li and Yong-Hui Jiang \\
School of Artificial Intelligence, Hainan Normal University, Haikou 571158, China}
\title{CDIO-CT collaborative  strategy for solving complex STEM problems in system modeling and simulation: an illustration of solving the period of mathematical pendulum}
\date{Dec. 28, 2025}

\begin{document}
\maketitle

\begin{abstract}
The problem-project-oriented STEM education plays a significant role in training students' ability of innovation. Although the conceive-design-implement-operate (CDIO) approach and the computational thinking (CT) are hot topics in recent decade, there are still two deficiencies: the CDIO approach and CT are discussed separately and a general framework of coping with complex STEM problems in system modeling and simulation is missing. In this paper, a collaborative strategy based on the CDIO and CT is proposed for solving complex STEM problems in system modeling and simulation with a general framework, in which the CDIO is about ``how to do", CT is about ``how to think", and the project means ``what to do".  As an illustration, the problem of solving the period of mathematical pendulum (MP) is discussed in detail. The most challenging task involved in the problem is to compute the complete elliptic integral of the first kind (CEI-1). In the  philosophy of STEM education, all problems have more than one solutions. For computing the CEI-1, four methods are discussed with a top-down strategy, which includes the infinite series method, arithmetic-geometric mean (AGM) method, Gauss-Chebyshev method and Gauss-Legendre method. The algorithms involved  can be utilized for R \& D projects of interest and be reused according to the requirements encountered. The general framework for solving complex STEM problem in system modeling and simulation is worth recommending to the college students and instructors.  
\end{abstract}

\tableofcontents

\section{Introduction}  \label{sec-Intro}
STEM education refers to the integration of Science, Technology, Engineering and Mathematics into the teaching and learning process, which stimulates massive innovation activities in the worldwide. 
STEM emphasizes that students should collaboratively observe, operate, experience and explore in the study process, and finally learn to create, so as to systematically improve their learning ability and problem-solving ability \cite{Straker2014,STEM-Map-2.0,Hafez2019}. 
The philosophy of STEM education is that each problem has more than one way to the feasible solutions. 

There are two dimensions for the teaching method of STEM education: one is the  problem which concerns ``what to do'', another is the project which concerns ``how to do''. In other words, the STEM education is problem-project-oriented.  The teaching strategy of the problem-project-oriented STEM education is the \textit{project-driven teaching} (PDT)\cite{Katz1989PDT,Li2017PDT,Peng2023PDT}. The PDT is also named with \textit{problem-based learning} (PBL1) \cite{Albabese1993PBL,Savery1995PBL,Wood2008PBL,Sabapathy2020ProjBL}  and \textit{project-based learning} (PBL2) \cite{Turnbull1999ProjBL,Thomas2000ProjBL,Li2017ProjBL,Saad2022ProjBL-CT}. Here the "learning" emphasizes the perspective of the students  and the "teaching" emphasizes the perspective of instructors/teachers. 

\begin{figure}[htbp]
\centering
\includegraphics[width=0.8\textwidth]{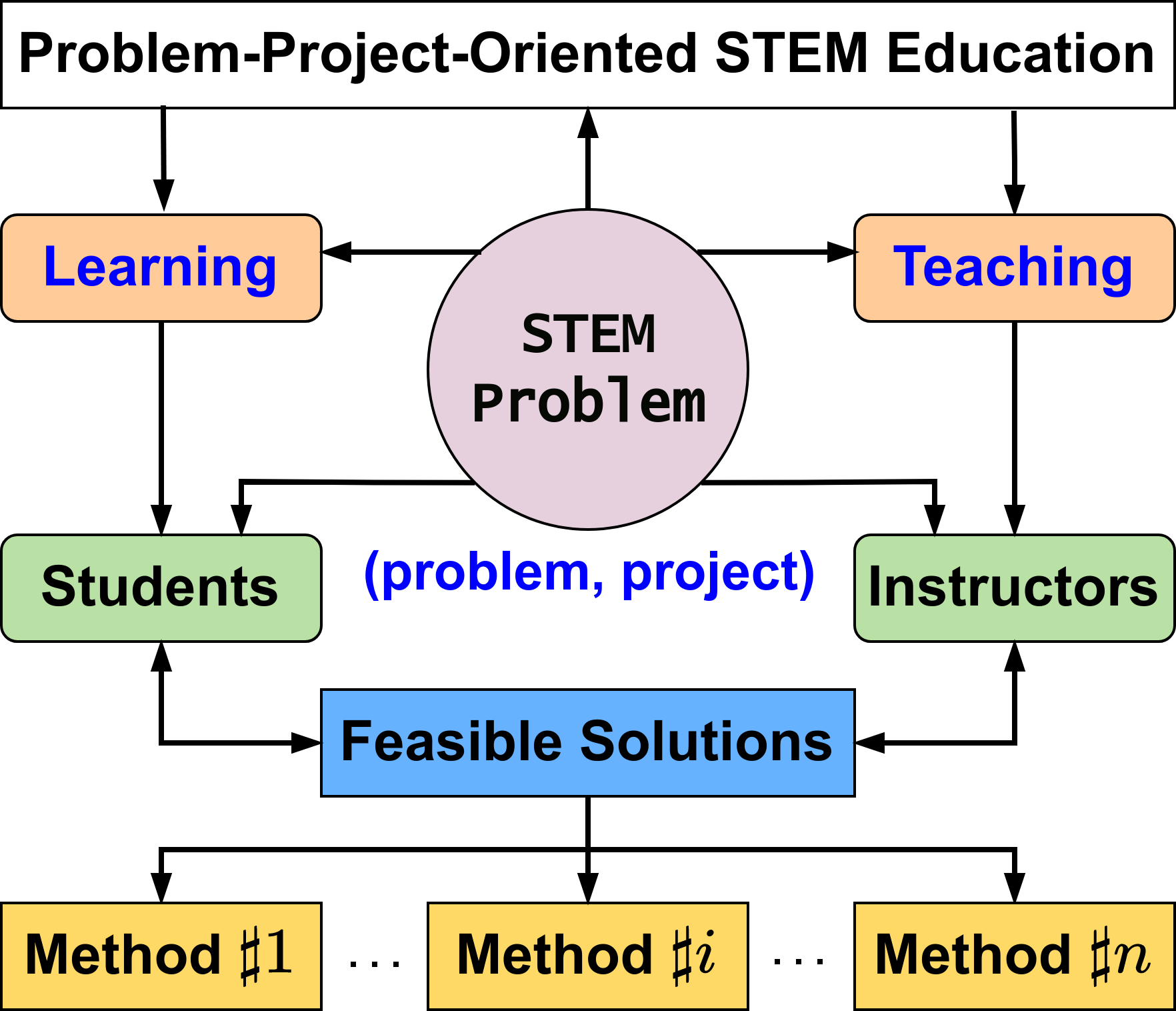} 
\caption{Problem-Project-Oriented STEM education}
\label{fig-PPO-stem}
\end{figure}

It is a pity that a unified terminology is missing for the teaching method of STEM education.  
In medical education, the problem-based learning is popular. In engineering education, the project-based learning and project-driven teaching are also popular. In K12 education, the project-driven teaching is accepted. It seems that the \textit{problem-project-oriented learning} (PPOL) might be a common name for them.  \Fig \ref{fig-PPO-stem} demonstrates the structure of problem-project-oriented STEM education emphasizing both dimensions of the teaching method. The feature of tackling one problem with multiple solutions results in a natural way of dividing the students into teams for different projects and shows the multidisciplinary properties of STEM education clearly.  

As the indication of E for engineering in STEM, the STEM education is closely related with the engineering education whose purpose is to train students' ability of \textit{solving complex engineering and technology problems} with the  well-known CDIO approach \cite{CDIO2014}. The CDIO stands for conceiving, designing, implementation and operation. It takes the life cycle from product R \& D to product operation as the carrier to enable students to learn engineering in an active, practical and organic way.  CDIO training outline divides the abilities of engineering graduates into four levels \cite{CDIO2014,SyllabusCDIOv3}: basic engineering knowledge, personal ability, interpersonal team ability and engineering system ability. The outline requires a comprehensive training method to enable students to achieve the predetermined goals at these four levels.

The STEM education is also closely connected with thinking methods. In 2006, Jeannette M. Wing \cite{Wing2006CT} proposed the concept of \textit{computational thinking} (CT). It has been approved as a fundamental ability which is parallel to the reading, writing and arithmetic  abilities in the traditional education concept in recent twenty years. The CT consists of five essential parts \cite{Wing2006CT,Voskoglou2012CT,Shute2017CT,Beecher2017CT,Intro2CT-MIT2020}: 
problem-solving and decomposition, 
abstraction and modeling, 
logical and algorithmic thinking,  anticipating and debugging (handling errors), and
\textit{verification-validation-testing} (VVT).
The CT integrates three kinds of thinking methods into a novel method, which includes
\begin{itemize}
	\item  the general mathematical thinking method used for problem solving, 
	\item the general engineering thinking method for the design and evaluation of huge and complex systems in the real world, and 
	\item the general scientific thinking method for understanding  complexity, intelligence, psychology and human behavior.
\end{itemize}
Although the STEM education, the CDIO approach and the CT are hot topics and 
there are quite a number of contributions dealing with them 
in recent decade, there are still some deficiencies:
\begin{itemize}
\item The CDIO approach and CT are discussed separately, thus their relation and interactions are ignored. For example, Sanford and Naidu \cite{Sanford2017} considered the relation of mathematical modeling and CT; Qian and Choi \cite{Qian2023} studied the process of developing abstraction in CT; 
Tan et al. \cite{Tan2023} discussed the balancing disciplinary knowledge and practical reasoning in problem solving; Nadelson et al. \cite{Nadelson2017} exhibited the STEM spectrum in K12 education. Ju\v{s}kevi\v{c}ien\.{e}  et al. \cite{Juskevicien2020CT} presented the design process of the strategy for CT abilities development in basic school. However, the relation and interaction of the CDIO approach and CT are not considered in these studies.
\item There is a lack of general framework of coping with complex STEM problem in system modeling and simulation.
 Mertens \cite{Mertens1977ModelSimu} proposed an earlier method for modeling and simulation, which includes
organizing information and crystallizing the thinking, 
identifying new research and teaching areas and techniques, and 
testing research and hypotheses. 
Mertens also presented nine guidelines for  the teachers or researchers to develop and utilize these models. Although Mertens' study provided valuable information for how to solve complex STEM problems, it is not systematically developed in the past 40 years. 
Magana and  Coutinho \cite{Magana2016} proposed a framework for identifying different audiences of computing and related CT practices at the intersection of computer science and engineering. Magana and Jong \cite{Magana2018} not only proposed approaches  for modeling-and-simulation-centric course design, teaching practices and pedagogy for modeling and simulation implementation, but also presented the evidence of learning with and about modeling and simulation practices. 
In these studies, the CT is deeply developed and the framework is touched, but the role of CDIO approach is missing. 
Mora et al \cite{Mora2015} and Zourmpakis et al \cite{Zourmpakis2023} pointed out that a clear and formal design process is important and an appropriate framework is necessary for the success of PDT. However, this study is not extended to solve complex STEM problems.
\end{itemize}
Actually, there are some interleaved questions to be considered deeply: 
\begin{itemize}
\item what is the relationship between STEM problem about ``what to do'', CDIO approach about ``how to do'' and CT about ``how to think'' in STEM education?  
\item how can  we merge the core ideas of STEM education, CDIO approach and CT into a new practical strategy to train college students via solving complex STEM problems? 
\item how can we understand and demonstrate the interaction  of CDIO approach and CT? 
\item how can we illustrate the STEM philosophy of tackling one problem with multiple ways to the solution?
\item how can we train teams of students in order to work on  multiple projects for solving the same problem in different ways?
\end{itemize}

\begin{figure*}[htbp]
	\centering
	\includegraphics[width=\textwidth]{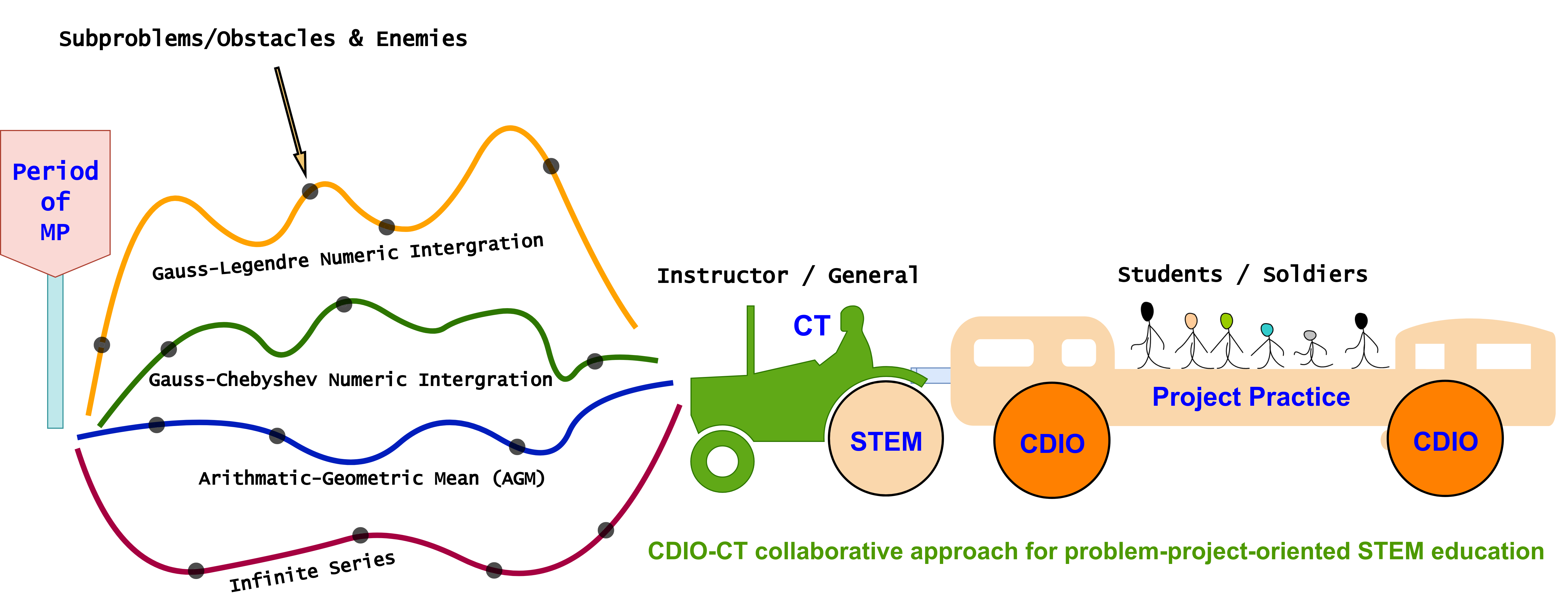}
	\caption{Metaphor of CDIO-CT collaborative strategy for 
	problem-project-oriented STEM education}
	\label{fig-road-map}
\end{figure*}

In order to answer these questions, we propose a strategy for the problem-project-oriented STEM education by combining the CDIO approach and CT effectively. In order to state the strategy concretely but not abstractly, we take the problem of solving  the period of \textit{mathematical pendulum} (MP) as an example for an illustration.
As a metaphor, \Fig \ref{fig-road-map} demonstrates the road map and essential ideas of our work intuitively. An effective way for training special forces via killing enemies to conquer objective position is the approach for training the students by solving the complex problem.
Figuratively speaking, the general's leading of the soldiers to conquer the objective position is similar to the instructor's guiding of the students to solve the objective problem, and the enemies in the battle are just like the sub-problems to be decomposed and solved during  working on the project. 

\begin{figure*}[htb]
\centering
\includegraphics[width=\textwidth]{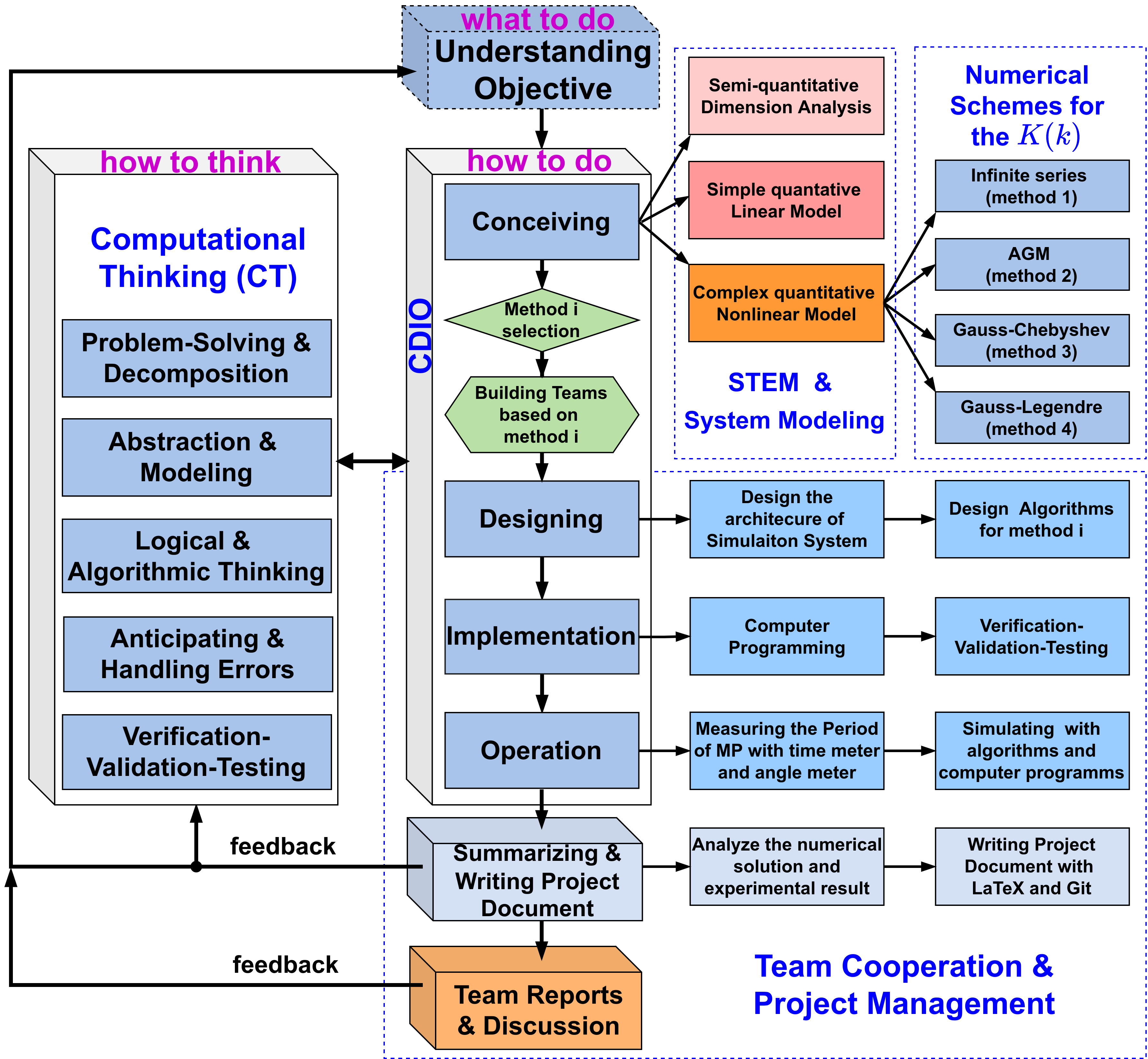} 
\caption{Practice of CDIO-CT collaborative strategy for STEM education by solving the period of MP}
\label{fig-team-coop}
\end{figure*}

In this paper, our main contributions lie in the following perspectives:
\begin{itemize}
\item An effective framework for solving complex STEM problem with the CDIO-CT collaborative strategy is proposed with the illustration of   a concrete problem --- solving the period of MP. 
The effective framework is shown in the block diagram \Fig \ref{fig-team-coop}, and it can be generalized to the general framework for solving complex STEM problems. 
\item The interaction of CDIO approach and CT is explored and explained with the help of concrete tasks in the project.
\item The philosophy of STEM education is demonstrated according to the dimensions of breath and depth:
    \begin{itemize}
    \item three ways for system modeling, viz. semi-quantitative method, simple quantitative method and complex quantitative method 
    are introduced respectively and comparatively so as to cope with the complexity of complex STEM problem and examine the performance of feasible solutions;
    \item four methods for computing the \textit{Complete elliptic integral of the first kind} (CEI-1) are discussed with the CDIO-CT collaborative strategy.
    \end{itemize}
\end{itemize} 

The rest contents of this paper are organized as follows: Section 
\ref{sec-methods} deals with the methods, which includes four parts: identifying the problem, solving the objective problem with CDIO approach, integration of CT and CDIO approach and PPOL for solving the period of MP;  Section \ref{sec-result} copes with the results; Section \ref{sec-discussion} gives the discussion for the details of techniques and limitation of the general framework; Section \ref{sec-conclusions} is the summary for this paper.

For the convenience of reading, we list the abbreviations appeared in this paper in \Tab \ref{tab-jargons}.
\begin{table}[htb]
\centering
\caption{Nomenclatures} 
\label{tab-jargons}
\resizebox{\textwidth}{!}{
\begin{tabular}{cl}
\hline
\textbf{Abbreviation} & \textbf{Interpretation} \\
\hline 
 STEM & science-technology-engineering-mathematics \\
 CDIO & conceive-design-implement-operate\\
 CT   & computational thinking \\
 PDT  & project-driven teaching \\
 PBL  & problem-based learning (PBL1), project-based learning (PBL2) \\
 PPOL & problem-project-oriented learning\\
 MP   & mathematical pendulum, also named by simple pendulum \\
 CEI-1 & $K(k)$, complete elliptic integral of the first kind \\
 VVT   & verification-validation-testing \\
 AGM   & arithmetic-geometric mean \\
 IS-CEI-1 & infinite series method for computing CEI-1 \\
 AGM-CEI-1 & AGM method for computing CEI-1 \\
 GC-CEI-1  & Gauss-Chebyshev method for computing CEI-1 \\
 GL-CEI-1  & Gauss-Legendre method for computing CEI-1 \\
 ROP & ratio of period \\
\hline 
\end{tabular}
}
\end{table}

\section{Methods} \label{sec-methods}

\subsection{Identifying the Problem of Interest}

\subsubsection{Problem --- Solving the Period of MP}

\textit{Mathematical pendulum} , also named simple pendulum, is a classic example in physics \cite{BerkleyPhys1-1973,FeynmanPhys1-2005,Halliday2018} for both college students and senior middle school students.  

\begin{figure}[htp]
	\centering
	\includegraphics[width=0.7\textwidth]{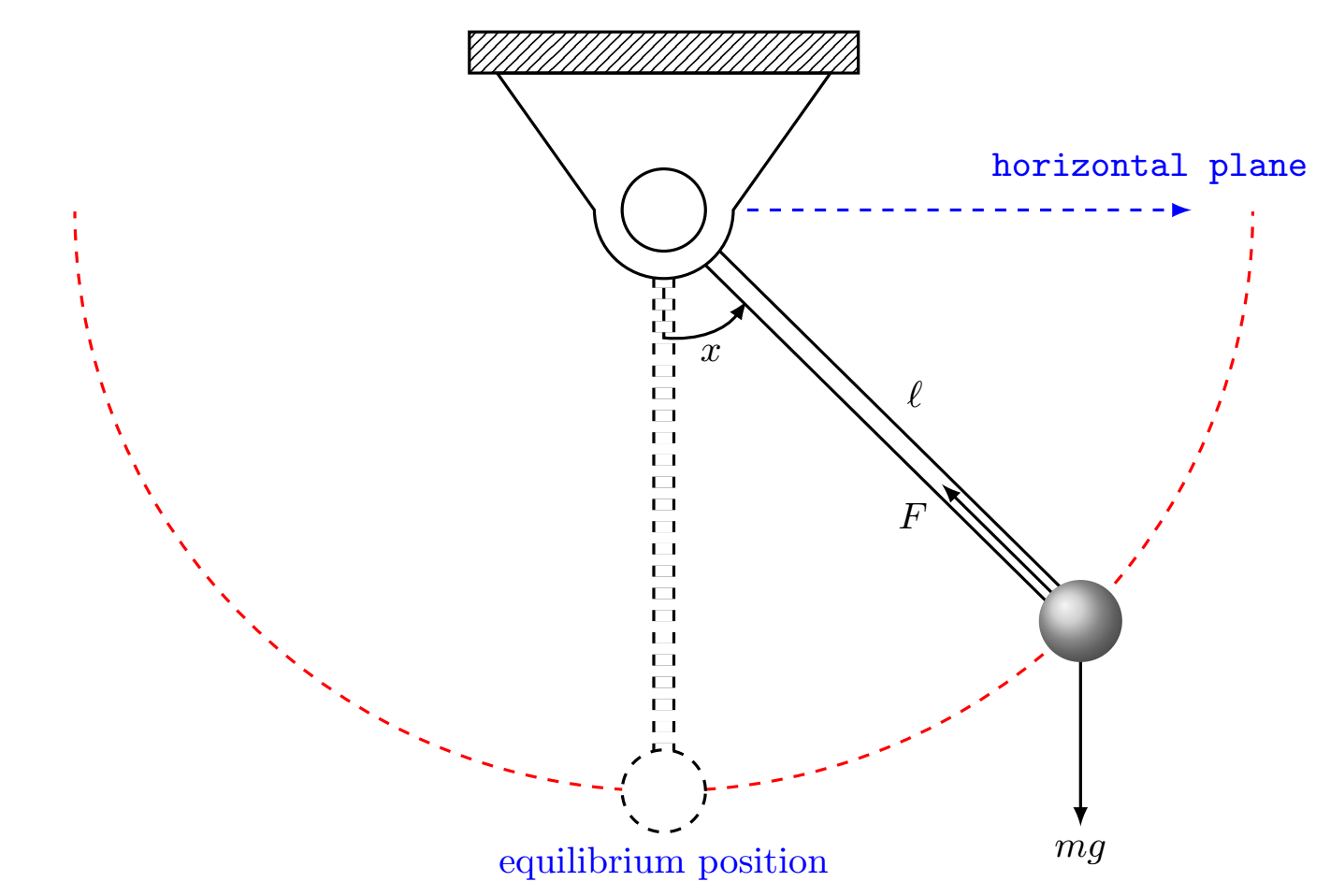} 
	\caption{Mathematical pendulum} \label{fig-MP}
\end{figure}

The mathematics of pendulums are in general quite complicated. Simplified assumptions can be made, which in the case of a MP allow the equations of motion to be solved analytically or numerically for periodic oscillations. A  \textit{mathematical/simple pendulum} is an ideal model of a \textit{real pendulum} subject to the following assumptions:
\begin{itemize}
	\item the bob is a point mass whose shape is ignored;
	\item the bob is fixed on a massless rod or cord, which is inextensible and always remains tight.
	\item the motion occurs only in two dimensions, i.e. the trajectory of the bob is an arc in a vertical plane, which is a closed manifold;
	\item the motion does not lose energy to friction or air resistance, thus the total energy, i.e., the sum of potential energy and kinematic energy is constant.
\end{itemize}
\Fig \ref{fig-MP} illustrates the idealization of the MP in which $\ell$ is the length of
rod, $m$ is the mass of the bob, $g$ is the acceleration of gravitation, and $x$ is the angular displacement. Theoretically, the amplitude of angular displacement $A$ measured in radius can be any positive value in the interval $(0, \frac{\pi}{2}]$. If $A = \pi/2$, then the trajectory of the MP will be half a circle instead of a small segment of arc. 

The mathematical model of the MP is described by the following motion equation
\begin{equation} \label{eq-MP-nonlinear}
	\secode{x}{t} + \Omega^2 \sin x = 0, \quad -A\le x \le A
\end{equation}
where
\begin{equation}
	\Omega = \sqrt{\frac{g}{\ell}}
\end{equation}
is the parameter of angular frequency. The problem discussed in this paper is:
\begin{StemProb}
Given the angular amplitude $A\in (0, \frac{\pi}{2})$, calculate the period of the MP effectively. 
 \end{StemProb}
We give some supplementary interpretations here:
\begin{itemize}
\item[a)] For any maximum angular displacement $A$,  we can find general and essential result for the relation of the period of MP with the parameters $A$, $\ell$, $m$ and $g$ illustrated in \Fig \ref{fig-MP}. 
\item[b)] The maximum angular displacement $A$ may be small such that $\sin x \sim x$ for any $x\in [-A, A]$, thus \eqref{eq-MP-nonlinear} can be approximated by a linear ODE
 \begin{equation} \label{eq-MP-linear}
	\secode{x}{t} + \Omega^2  x = 0, \quad -A\le x \le A.
\end{equation}
\item[c)] The maximum angular displacement $A$ may be large such that we can not use \eqref{eq-MP-linear} and we have to cope with the nonlinear ODE \eqref{eq-MP-nonlinear}.  
\end{itemize} 

\subsubsection{Exploring Period Formula via System Modeling}

There are three fundamental schemes for finding the period of MP, which demonstrates different ways for solving the period.  

\subsubsection*{A. Semi-Quantitative Method via Dimension Analysis}

We can find the semi-quantitative result of period with the dimension analysis \cite{Polya1957,HandbookMPES2011}. Actually, let the dimensions for the length, mass and time are $\physdim{L}, \physdim{M}$ and $\physdim{T}$ respectively, then the dimension for the gravitation acceleration $g$ will be $[g] = \physdim{L}\physdim{T}^{-2}$. Assume the dimension of period be 
\begin{equation}
	[T] = [m]^\alpha [g]^\beta [\ell]^\gamma,
\end{equation}
then according to the definition of dimension of physics quantities, we have
\begin{equation}
	\physdim{T} = \physdim{M}^\alpha \cdot \physdim{L}^\beta\physdim{T}^{-2\beta} 
	\cdot \physdim{L}^\gamma = \physdim{M}^\alpha  \physdim{L}^{\beta +\gamma} \physdim{T}^{-2\beta}. 
\end{equation}
The equilibrium of dimensions shows that
\begin{equation}
	\left\{
	\begin{split}
		\alpha &= 0 \\  \beta + \gamma &= 0 \\ -2\beta &= 1 \\
	\end{split} 
	\right. \Longleftrightarrow 
	\begin{bmatrix} 1 & 0 & 0\\ 0 & 1 & 1 \\ 0 & -2 & 0 \end{bmatrix}
	\begin{bmatrix} \alpha \\ \beta \\ \gamma \end{bmatrix}
	=\begin{bmatrix}
	0 \\ 0 \\ 1
	\end{bmatrix}.
\end{equation}
Consequently, $\alpha = 0, \beta = -1/2, \gamma = 1/2$, which implies that
\begin{equation} \label{eq-T-lambda}
	T = \lambda \sqrt{\frac{\ell}{g}}  \propto \sqrt{\frac{\ell}{g}},
\end{equation}
where the unknown constant coefficient $\lambda$ can be determined empirically with observation data collected in practical experiments of MP. 

We remark that the generalization of dimension analysis  is the \textit{type checking} 
in the sense of CT \cite{Wing2006CT}. Without doubt, for the same type of physics quantity, they must have the same unit and dimension. 

\subsubsection*{B. Simple Quantitative Method based on Linear Approximation}

If the amplitude $A$ for the angular displacement is sufficiently small, say $ A < \pi/36 ~\mathrm{rad}$ 
(or $ A$ is less than $5^\circ$ equivalently), then the linear ODE \eqref{eq-MP-linear} implies  the famous formulae of period
\begin{equation} \label{eq-T-linear}
	\scrd{T}{linear}  = \frac{2\pi}{\Omega} = 2\pi \sqrt{\frac{\ell}{g}}.
\end{equation}  
In this linear approximation, we find that 
\begin{equation}
	\scrd{\lambda}{linear}  = 2\pi
\end{equation}
in the sense of small amplitude of angular displacement.

\subsubsection*{C. Complex Quantitative Method via General Nonlinear Model}

\begin{figure*}[htb]
	\centering
	\includegraphics[width=0.9\textwidth]{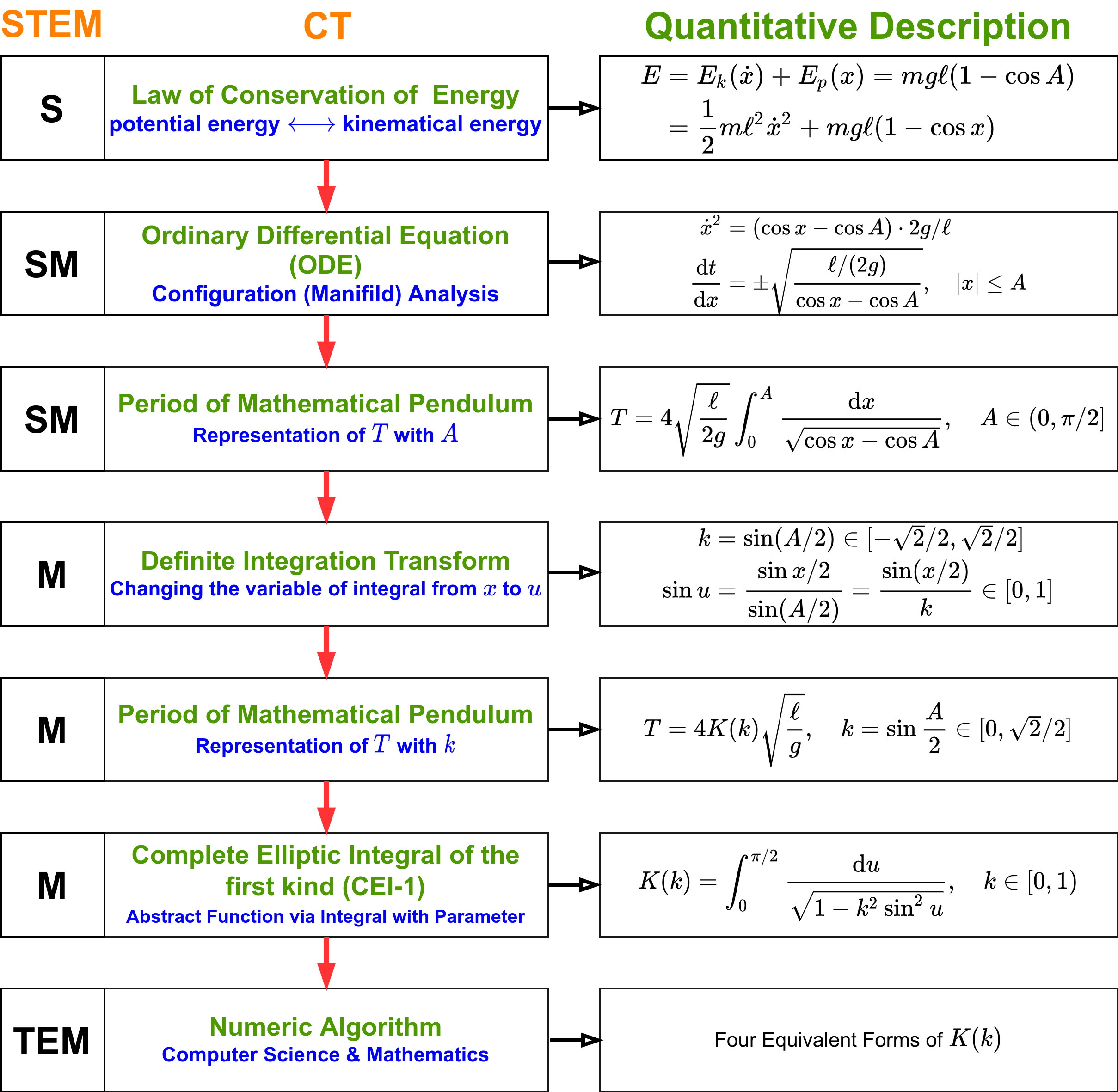} 
	\caption{Identifying the essential STEM problem of solving the period of Mathematical Pendulum} 
	\label{fig-MP-STEM}
\end{figure*}

Solving the period of MP with an arbitrary angular amplitude is a typical complex problem which integrates the four components of STEM.  \Fig \ref{fig-MP-STEM} outlines the procedures  to derive the formula of period in a step-by-step way.
For the angular displacement $x\in [-A, A]$, let
\begin{equation}
	k = \sin \frac{A}{2}, \quad \sin u = \dfrac{\sin \dfrac{x}{2}}{\sin \dfrac{A}{2}},
\end{equation}
with the help of the conservative law of energy and some mathematical techniques, we can find the following general formula of the MP
\begin{equation} \label{eq-T-eci1}
		T = 4 K(k)\cdot \sqrt{\frac{\ell}{g}}, \quad k\in \left[0, \frac{\sqrt{2}}{2} \right]
\end{equation}
where 
\begin{equation} \label{eq-K-ku}
K(k) = \int^{\pi/2}_0 \frac{\dif u}{\sqrt{1-k^2 \sin^2 u}}, \quad k\in [0,1]
\end{equation}
is the CEI-1 \cite{TableISP2015,Abramo1965}. 
A comparison of \eqref{eq-T-lambda}, \eqref{eq-T-linear} and \eqref{eq-T-eci1} implies that 
\begin{equation}
\scrd{\lambda}{nonlinear} = 4K(k), \quad 
\scrd{\lambda}{linear} = 4K(0) = 2\pi.
\end{equation}

\begin{table}[ht]
	\centering
	\caption{Expression of period $\displaystyle T = \lambda \sqrt{\frac{\ell}{g}}$}
	\label{tab-T-lambda}
	\begin{tabular}{ll}
		\hline
		\textbf{Method} & \textbf{Coefficient} $\lambda$ \\
		\hline
		Dimension analysis method & $\lambda$ is an unknown constant \\
		Simple quantitative method & $\scrd{\lambda}{linear} = 2\pi= 4K(0)$ \\
		Complex quantitative method & $\scrd{\lambda}{nonlinear} = 4K(k)$\\
		\hline 
	\end{tabular}
\end{table}
\Tab \ref{tab-T-lambda} shows the expressions of the period for different analysis methods. It is clear that the more reasonable the model is, the more accurate the expression of period is.  Obviously, we can find that: the dimension analysis method gives the general relation of $T$, $\ell$, $m$ and $g$ but leaves the coefficient $\lambda$ unspecified; the complex quantitative method returns the finest result; the simple quantitative method returns a relatively coarse result which connects the general and finest results.

\subsubsection{Identifying the STEM Problems }

Solving the period of MP with an arbitrary angular amplitude is a typical complex problem which integrates the four components of STEM.  Once the CEI-1 is obtained, the ultimate problem left will be the computation of $K(k)$, which is a combination of mathematics and engineering. Here, the exact meaning of engineering refers to computer programming and software engineering. In other words, for the general nonlinear model of the motion equation of MP, the most challenging task in determining the period of MP is equivalent to computing the value of the special function $K(k)$.  

\begin{figure}[h]
	\centering
	\includegraphics[width=0.8\textwidth]{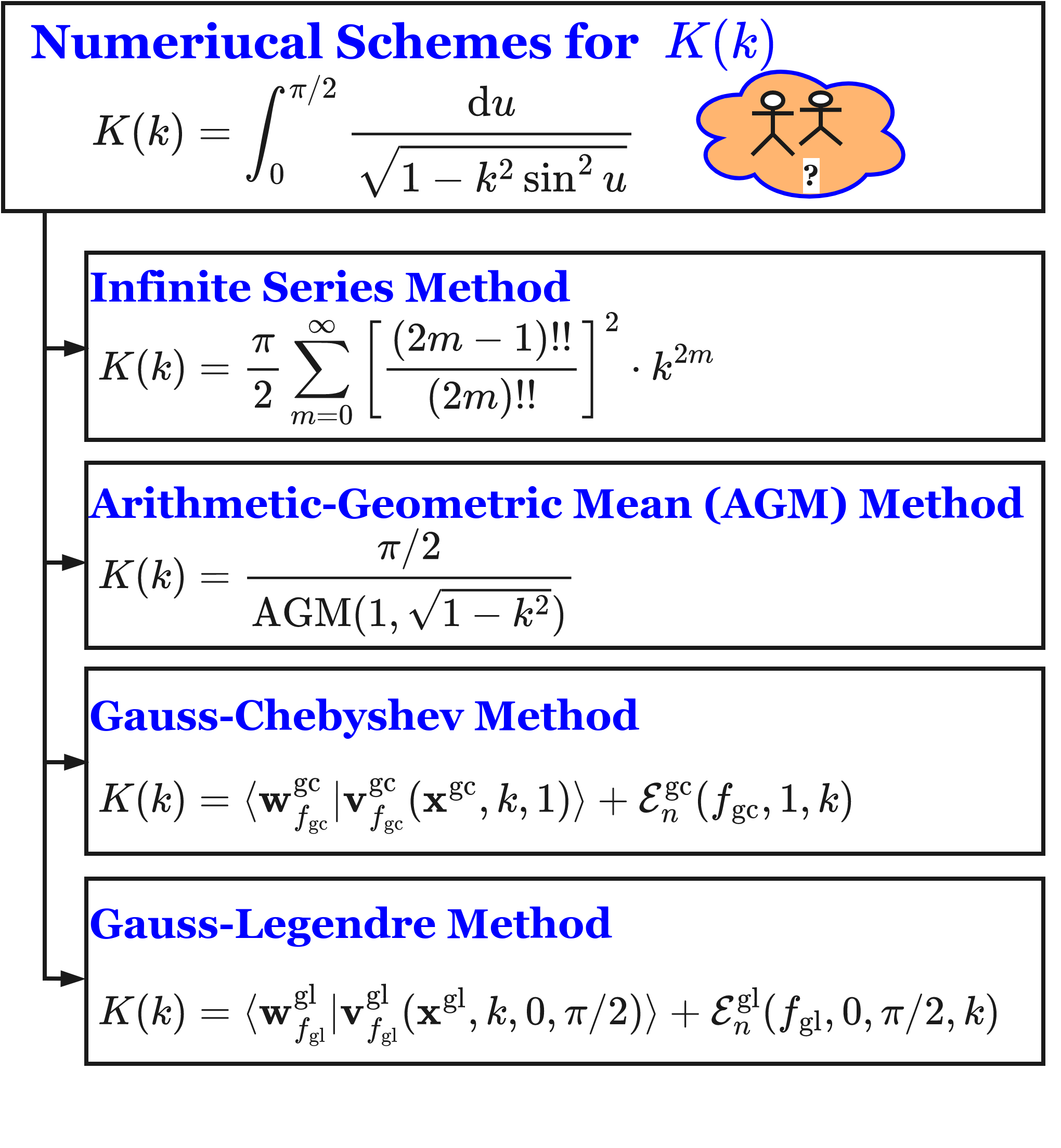} 
	\caption{Four methods for computing CEI-1}
	\label{fig-ways2K}
\end{figure}

In this paper, we will discuss four methods to calculate the special function $K(k)$ as shown in \Fig  \ref{fig-ways2K}.

\subsection{Solving Problem with CDIO Approach}

\subsubsection{Conceiving}

It is difficult for us to calculate the value of CEI-1 directly since there is no available simple formulae for this kind of integral with extra parameter $k$. The purpose of conceiving is to explore the way to express the CEI-1 with reasonable approximation or equivalent form which can be computed conveniently with computer program such that the precision is good enough and the computational complexity is acceptable. Fortunately, we can find three fundamental kinds of methods to do so:
\begin{itemize}
	\item Expanding the CEI-1 into an infinite series which converges rapidly;
	\item Computing the CEI-1 with the arithmetic-geometric mean (AGM) method which converges rapidly and iterates simply;
	\item Converting the CEI-1 into a finite sum with $2n-1$ order algebraic precision by taking the $n$-th order orthogonal polynomials via Gauss numerical integration method. 
\end{itemize}	
The original Gauss numerical method can be expressed by	
\begin{equation}
I(f) = \int^b_a \rho(x)f(x)\dif x 
	= \sum^n_{i=1}w_if(x_i) + \mathcal{E}_n(f)
\end{equation}
	where $n$ is a positive integer, $\set{x_i: 1\le i \le n}$ are the $n$ roots of an orthogonal polynomial, say $ Y_n(x)$, $\set{w_i: 1\le i \le n}$ are the weights specified by the roots of $Y_n(x)$, and $\mathcal{E}_n(f)$ is the error of optimal approximation which will be omitted according to the requirement of precision in practice. 
	However, the integral $K(k)$ is more complicated since it is defined by an integral with extra parameter $k$. With the help of the concepts of abstract function defined by an integral and developed in functional analysis, we can use the following form	
\begin{equation}
\begin{split}
K(k) &=\int^b_a f(x,k)\dif \mu 
= \int^b_a f(x,k)\rho(x)\dif x \\
&= \sum^n_{i=1} w_i f(x_i, k) + \mathcal{E}_n(f)
\end{split}
\end{equation}
where $\rho(x)$ is the non-negative weight function and the measure $\mu$ is specified by $\dif \mu = \rho(x)\dif x$.
	There are two feasible methods for computing $K(k)$ in the sense of Gauss numeric integration:
	\begin{itemize}
		\item Gauss-Chebyshev method, where $\displaystyle \rho(x) = \frac{1}{\sqrt{1-x^2}}$ and $b = -a = 1$;
		\item Gauss-Legendre method, where $\rho(x) = 1, b = \pi/2$ and $a = 0$.
	\end{itemize}    

As shown in \Fig\ref{fig-ways2K}, there are four  key formula for computing the ECI-1 $K(k)$ as follows: 
\begin{itemize}
\item Infinite series method: the $K(k)$ can be computed with the following infinite series 
\begin{equation} \label{eq-K-series}
K(k) = \frac{\pi}{2} \cdot  \sum^\infty_{m=0}\left[\frac{(2m-1)!!}{(2m)!!} \right]^2k^{2m} 
\end{equation}
\item AGM method: the relation of $K(k)$ and AGM can be expressed  by
\begin{equation} \label{eg-K-agm}
K(k) = \frac{\pi/2}{\AGM(1, \sqrt{1-k^2})}
\end{equation}
\item Gauss-Chebyshev method: the $K(k)$ is a special case of the the abstract function defined by
\begin{equation}
	\scrd{F}{C}(k)  = \int^a_{-a} \frac{f(x, k)}{\sqrt{a^2 - x^2}}\dif x 
	= \int^1_{-1} \frac{f(at, k)}{\sqrt{1-t^2}}\dif t
\end{equation}
which  can be computed by the extended Gauss-Chebyshev formulae of numerical integration. Actually, we can deduce that \cite{Zhang2022-CEI1}
\begin{equation}\label{eq-K-gc}
			K(k) =\braket{\scru{\vec{w}}{gc}}{\scru{\vec{v}}{gc}_{\scrd{f}{gc}}} + 
			\scru{\mathcal{E}}{gc}_n
\end{equation}
where  $\scru{\mathcal{E}}{gc}_n$  is the negligible approximation error of numerical integration and 
\begin{equation}
	\left\{
	\begin{split}
		&\scru{\vec{x}}{gc} = \trsp{[\scru{x}{gc}_1, \cdots, \scru{x}{gc}_n]},\\
		&\scru{\vec{w}}{gc} = \trsp{[\scru{w}{gc}_1, \cdots, \scru{w}{gc}_n]},\\
		&\scru{\vec{v}}{gc}_{\scrd{f}{gc}} = \trsp{[\scrd{f}{gc}(\scru{x}{gc}_1,k), \cdots, \scrd{f}{gc}(\scru{x}{gc}_n,k)]}
	\end{split}
	\right.
\end{equation}
in which $(\scru{x}{gc}_i, \scru{w}{gc}_i)$ are the root-weight pairs such that
\begin{equation} \label{eq-x-w-gc}  
\scru{x}{gc}_i = \cos \frac{(2i-1)\pi}{2n}, \quad 
	\scru{w}{gc}_i = \frac{\pi}{n}, \quad 1\le i\le n
\end{equation}
and 
\begin{equation} \label{eq-f-gc}  
f(t,k) = \scrd{f}{gc}(t,k) =  \frac{1/2}{\sqrt{1-k^2t^2}}.
\end{equation}
\item Gauss-Legendre method:  the $K(k)$ is a special case of the the abstract function defined by
\begin{equation}
\begin{split}
		\scrd{F}{L}(k)
		&=\int^b_a f(x,k)\dif x  \\
		&= \frac{b-a}{2}\int^1_{-1} f\left(\frac{b+a}{2}+\frac{b-a}{2}t \right)\dif t
		\end{split}
\end{equation}
which  can be computed 
by the extended Gauss-Legendre formulae of numerical integration. 
Let 
$$a = 0, b = \pi/2$$
and 
\begin{equation} \label{eq-f-gl}
	f(u, k) = \scrd{f}{gl}(u,k) =  \frac{1}{\sqrt{1-k^2\sin^2u}},
\end{equation}
we can deduce that \cite{Zhang2022-CEI1}
\begin{equation} \label{eq-K-gl}
		\begin{split}
			K(k) &= \frac{\pi}{4}\sum^n_{i=1} \scru{w}{gl}_i \scrd{f}{gl}\left(\frac{\pi}{4}(1+\scru{x}{gl}_i), k\right) 
			+  \scru{\mathcal{E}}{gl}_n \\
			&= \braket{\scru{\vec{w}}{gl}}{\scru{\vec{v}}{gl}_{\scrd{f}{gl}}} +   \scru{\mathcal{E}}{gl}_n
		\end{split}
\end{equation}
where 
\begin{equation}
\scru{\vec{x}}{gl}  = \trsp{\left[\scru{x}{gl}_1, \cdots, \scru{x}{gl}_n\right]}, \scru{\vec{w}}{gl}  = \trsp{\left[\scru{w}{gl}_1, \cdots, \scru{w}{gl}_n\right]}
\end{equation}
are the root and weight vectors such that 
\begin{equation} \label{eq-weight-Legendre}
\left\{
\begin{split}
& P_n(\scru{x}{gl}_i) = 0, \quad 1\le i \le n \\
& \scru{w}{gl}_i 
= \frac{2}{n}\cdot \frac{1}{P_{n-1}(\scru{x}{gl}_i)P'_n(\scru{x}{gl}_i)}
\end{split}
\right.
\end{equation}
for the $n$-th order Legenedre polynomial $P_n(x)$ and its derivative $P'_n(x)$. Moreover, $\scru{\mathcal{E}}{gl}_n$ is the negligible approximation error of numerical integral and
\begin{equation}
\scru{\vec{v}}{gl}_{\scrd{f}{gl}}  = \frac{\pi}{4}\trsp{\left[\scrd{f}{gl}(\scru{u}{gl}_i,k), \cdots, \scrd{f}{gl}(\scru{u}{gl}_n,k)\right]}
\end{equation}
in which
\begin{equation}
\scru{u}{gl}_i  = \frac{\pi(1 + \scru{x}{gl}_i)}{4}, \quad 1\le i\le n
\end{equation}
for sufficiently large $n$ when the approximation $\scru{\mathcal{E}}{gl}_n$ is given.
\end{itemize}

\subsubsection{Designing}

In the sense of CDIO educational approach, the objective of design is to propose feasible procedures according to the conceiving process for solving computation oriented problems concerned. In this paper, the purpose of design is to present  algorithms for computing CEI-1 efficiently, robustly and precisely based on the four quantitative methods discussed above. In the conceiving process, we first identify a general problem, then specify some parameters so as to establish its connection with CEI-1. Similarly, in the design process we  first design an algorithm for solving the general problem, then specify some parameters to get specific algorithm for computing CEI-1. For more details about the design of numerical algorithms for the CEI-1, please see the reference \cite{Zhang2022-CEI1}.

\subsubsection{Implementation}

In the sense of CDIO educational approach, the implementation process follows the design process directly. For the software-oriented tasks, the key issue of implementation is to present the feasible code for the algorithms with some specific computer programming language. Without loss of generality for the implementation process,  we choose the C programming language \cite{RK-C-1988} to implement our design for computing the  CEI-1 $K(k)$. In other words, we discuss how to write the C codes ``technologically" for computing the CEI-1 with four different methods and corresponding algorithms. For this purpose, we rely on a fundamental principle --- separation of interface and implementation details --- to simplify the implementation process, reuse the code for other problems and upgrade system in the future.

The principle of separating interface and implementation details, also named by information hiding, is frequently encountered in solving complex engineering and technology problems. There are some typical reasons for adopting this separation rule:
\begin{itemize}
	\item it is not necessary for the users to known the details of the implementation;
	\item the implementation is encapsulated to hide information about the details so as to
	\begin{itemize}
		\item simplify the system structure by integrating the implementation into a basic module as an element for complex system,
		\item provide simple user interface for system connection,
		\item replace the module with different implementation while keeping the same interface when updating/maintaining the system and
		\item reduce the safety risk by prohibiting the access to details of implementation;
	\end{itemize}
	\item the implementation is closely connected with the intellectual property rights which limits the access authority. 
\end{itemize}

The interfaces of algorithms consist of the declarations of various procedures involved in concrete computation tasks. These declarations can be organized as different header files and the corresponding details of implementation will be organized as different source files. According to the sub-problems decomposed in the CDIO process, the objective of solving CEI-1 can be conquered by finishing eight sub-problems and the corresponding modules of software.  \Tab \ref{tab-modules-comp-K} illustrates the modules, interface files and implementation files involved in computing CEI-1 with four different methods. Our design and implementation of the software are based on the C programming language. The code for the implementation can be downloaded from the following  GitHub site 
\begin{center}
\textcolor{blue}{\href{https://github.com/GrAbsRD/CEI-1/tree/main}{https://github.com/GrAbsRD/CEI-1/tree/main}}
\end{center}

The details of implementation of calculating CEI-1 consist of a series of source files, please see the file names \cpfile{*.c} in \Tab \ref{tab-modules-comp-K}. We remark that the file \cpfile{main.c} is used to generate a table of the values of CEI-1 for different $k$ computed with different methods, which can be compared with its counterpart appearing in the handbook about CEI-1.

\begin{table*}[htp]
\centering
	\caption{Modules, interfaces and dependencies for computing CEI-1} \label{tab-modules-comp-K}
	\begin{tabular}{p{1cm}p{6cm}p{5cm}p{3cm}}
		\hline
		\textbf{No.} & \textbf{Modules for CEI-1 Solver}  & \textbf{Interface File}   & \textbf{Dependencies}\\
		\hline
		\hline
		\ding{172} & Abstract Data Type &  \cpfile{DataType.h} &     \\
		\ding{173} &  Usual Mathematics &  \cpfile{UsualMath.h}    &  \ding{172} \\
		\ding{174} & Orthogonal Polynomial &  \cpfile{OrthoPolynomial.h}  & 
		\ding{172}\ding{173} \\
		\ding{175}   & Fixed Point &  \cpfile{FixedPoint.h}  & \ding{172}  \\
		\hline
		\ding{176} & Infinite Series Solver&  \cpfile{InfiniteSeries.h}   & \ding{172}\\
		\ding{177} & AGM Solver &  \cpfile{AGM.h}  & \ding{172}\ding{173} \\
		\ding{178} & Gauss-Chebyshev Integral Solver &  \cpfile{IntegralSolverGC.h}  & \ding{172}\ding{173}\\
		\ding{179} & Gauss-Legendre Integral Solver &  \cpfile{IntegralSolverGL.h}  & \ding{172}$\sim$\ding{175}\\
		\hline
		\ding{180} & Complete Elliptic Integral Solver & \cpfile{CEI.h}  & \ding{172}$\sim$\ding{179}\\
		\hline
		\checkmark & Verification-Validation-Testing & \cpfile{main.c} &   
		\ding{172}$\sim$\ding{180} \\
		\hline  
	\end{tabular}
\end{table*}

\subsubsection{Operation}

The results of numerical analysis of CEI-1 can be used for different scenarios such as 
\begin{itemize}
\item[1)] Application deployment in the experiment of measuring the period:
\item[2)] Applications of the methods and algorithms developed: The methods developed for solving the period can be used for many problems. There are lots of byproducts for quantitative analysis such as infinite series algorithm, AGM algorithm, Gauss-Chebyshev and Gauss-Legendre algorithm. The technique of solving fixed-point is of  significance for STEM problems. 
\item[3)] Exploring the evolution of $T = \lambda \sqrt{\ell/g}$  from the unknown constant in dimension analysis to the constant $2\pi$ in linear approximation and $4K(k)= 2\pi(1+ k^2/4 + 9k^4/64 + \cdots)$ in the nonlinear equation of the motion. 
The variation of the period with the angular amplitude means that it is not a good idea to use MP as a timer since the angular amplitude is difficult to be controlled precisely.

\end{itemize}

\subsection{Integration of CT and CDIO Approach}

\subsubsection{Problem-solving and Decomposition}

The fundamental method for solving complex engineering and technology problems is decomposing the problem into solvable sub-problems. Alternatively, a large complex problem is divided into smaller, simpler sub-problems. The motivation behind it is that solving a series of simple problems is easier than trying to tackle a big complex one. Decomposition is a basic technique to control complexity of the system concerned. 
In the sense of system theory, it means that we should take the following strategy: firstly, we decompose a system into a series of subsystems and define the interfaces logically and clearly; secondly, we build, test and validate each subsystem separately; thirdly, we integrate the subsystems into the final system; finally, we optimize the final system and take it into practice.

For each solving method for computing the CEI-1, there is a feasible   problem-decomposition solution to the computation task. 
\begin{itemize}
\item \Tab \ref{tab-PD-IS} illustrates the sub-problems of the infinite series solver for the CEI-1. The key issues lie in  binding the general infinite series and the concrete infinite series for the CEI-1 and calculating the coefficients $c_m$ efficiently and robustly.  

\begin{table}[htb]
\centering
\caption{Problem-decomposition of IS-CEI-1 solver}
\label{tab-PD-IS}
\begin{tabular}{cp{0.9\textwidth}}
\hline
\textbf{No}. & \textbf{Subproblems of the Infinite Series Solver for the CEI-1}  
\\
\hline \hline
0 & Pattern analysis --- binding the general infinite series $\displaystyle{\sum^\infty_{m=0}c_mx^m}$ and the CEI-1. \\
1   & Binding the parameter $x$ by assigning $x\leftarrow k^2$.\\
2   & Set the constant $\pi/2$ with a library function. \\
3   & Compute the infinite series $S(x) = \sum\limits^\infty_{m=0}c_mx^m$ iteratively with a loop. \\
    & 3.1 Compute the General Term $T_m(x) = c_m x^m$. \\
    & 3.2 Convergence Analysis: $\abs{T_m(x)}\overset{?}{\le}\varepsilon$.\\
\hline 
\end{tabular}
\end{table}

\item \Tab \ref{tab-PD-AGM} lists the sub-problems of the AGM solver for the CEI-1. The key problems include the connection of AGM algorithm and the CEI-1 proposed by King in \cite{King1924AGM} and the fixed-point algorithm for computing the value of $\mathrm{AGM}(a,b)$. 
\begin{table}[htb]
\centering
\caption{Problem-decomposition of AGM-CEI-1 solver}
\label{tab-PD-AGM}
\begin{tabular}{cp{0.9\textwidth}}
\hline
\textbf{No}. & \textbf{Sub-problems of the AGM Solver for the CEI-1 }  \\
\hline \hline
0 & Pattern analysis --- binding the AGM function $\AGM(a,b)$ and the  CEI-1. \\
1   & Binding the parameters $a, b$ by assigning $a \leftarrow 1$ and $b \leftarrow \sqrt{1-k^2}$.\\
2   & Set the constant $\pi/2$ with a library function. \\
3   & Compute the $\mathrm{AGM}(a, b, \varepsilon)$ with a fixed-point algorithm.\\
    & 3.1 Set the initial value $\ES{R}{2}{1}\ni\vec{x}_0 \leftarrow (a, b)$. \\
    & 3.2 Specifying the 2-dim updating function $\scrd{\mathcal{A}}{agm}$.\\
    & 3.3 Specifying the Euclidean distance function $\dist: \ES{R}{2}{1}\times \ES{R}{2}{1}\to \mathbb{R}^+$ for convergence decision.\\
    & 3.4 Compute the fixed-point $\vec{x}$ iteratively via $\vec{x}_{i+1} = \scrd{\mathcal{A}}{agm}(\vec{x}_i)$ with a fixed-point algorithm. \\  
\hline 
\end{tabular}
\end{table}

\item \Tab \ref{tab-PD-GC-CEI} demonstrates the sub-problems of the   Gauss-Chebyshev solver for the CEI-1. The essential difficulty lies in the conversion of $K(k)$ in \eqref{eq-K-ku} to the standard form of Gauss-Chebyshev numerical integration by changing the angle variable $u\in[0,\pi/2]$ to the parameter $t = \sin u \in [-1, 1]$.

\begin{table}[htb]
\centering
\caption{Problem-decomposition of GC-CEI-1 solver}
\label{tab-PD-GC-CEI}
\begin{tabular}{cp{0.9\textwidth}}
\hline
\textbf{No}. & \textbf{Sub-problems of the Gauss-Chebyshev  solver for the CEI-1}  \\
\hline \hline
0 & Pattern analysis --- binding the general integral $\displaystyle{\int^a_{-a} \frac{f(t,k)}{\sqrt{a^2-t^2}}\dif t}$ and the CEI-1. \\
1 & Specifying the integral kernel $\scrd{f}{gc}(t,k)$. \\
2 & Specifying the parameters number $n$ of nodes for Gauss numerical integral and the positive parameter $a$ for the symmetric interval $[-a, a]$.  \\
3 & Compute the integral $\displaystyle{\int^a_{-a} \frac{\scrd{f}{gc}(t,k)}{\sqrt{a^2 - t^2}}\dif t}$ by Gauss numerical integration method through inner product. \\
4 & Compute the real $n$-dim inner product $\braket{\scru{\vec{w}}{gc}}{\scru{\vec{v}}{gc}_f}: \ES{R}{n}{1}\times \ES{R}{n}{1}\to \Real $.\\
  &  4.1 Set the  weight vector $\scru{\vec{w}}{gc}$.\\
  &  4.2 Set the function value vector $\scru{\vec{v}}{gc}_{f}$ with the function $\scrd{f}{gc}$ and roots $\scru{\vec{x}}{gc}$ of the $n$-th order Chebyshev polynomial.\\ 
\hline   
\end{tabular}
\end{table}

\item \Tab \ref{tab-PD-GL-CEI} gives the sub-problems of the Gauss-Legendre solver for the CEI-1. There are three challenges: establishing the connection of $K(k)$ and the Gauss-Legendre numerical integration, generating Legendre polynomial $P_n(x)$ and its derivative $P'_n(x)$ efficiently, calculating the $n$ roots of $P_n(x)=0$ by Newton's method automatically.   

\begin{table}[htb]
\centering
\caption{Problem-decomposition of GL-CEI-1 solver}
\label{tab-PD-GL-CEI}
\begin{tabular}{cp{0.9\textwidth}}
\hline
\textbf{No}. & \textbf{Sub-problems of the Gauss-Legendre solver for the CEI-1}\\
\hline \hline
0 & Pattern analysis --- binding the general integral $\displaystyle{\int^b_a f(t,k)\dif t}$ and the CEI-1. \\
1 & Specifying the integral kernel $\scrd{f}{gl}(t,k)$. \\
2 & Specifying the parameters number $n$ of nodes for Gauss numerical integral and the parameters for the interval $[a, b]$.  \\
3 & Compute the integral $\displaystyle{\int^b_a f(t,k)\dif t}$ by Gauss numerical integration through inner product. \\
4 & Compute the $n$ roots of $P_n(x) = 0$. \\
  &  4.1 Compute the Legendre polynomials $P_n(x)$ iteratively. \\
  &  4.2 Compute the derivative of Legendre polynomials, i.e., the $P'_n(x)$. \\
  &  4.3 Compute the roots of nonlinear equation $f(x,\alpha) = 0$ with Newton's method via the iterative formulae $x_{i+1} = \scrd{\mathcal{A}}{Newton}(x_i,f,\alpha)=x_i - \frac{f(x_i,\alpha)}{f'(x_i,\alpha)}$.\\
  &  4.4 Specifying the initial values for computing the roots of $P_n(x) = 0$.\\
  &  4.5 Compute the roots of $P_n(x) = 0$ by  combining sub-problems 
    4.1 $\sim$ 4.4. \\        
5 & Compute the real $n$-dim inner product $\braket{\scru{\vec{w}}{gl}}{\scru{\vec{v}}{gl}_f}: \ES{R}{n}{1}\times \ES{R}{n}{1}\to \Real $\\
  &  5.1 Compute the  weight vector $\scru{\vec{w}}{gl}$ with the roots of $P_n(x) = 0$ and the derivative of $P'_n(x)$. \\
  &  5.2 Set the function value vector $\scru{\vec{v}}{gl}_{f}$ with the function $\scrd{f}{gl}$ and roots of the $P_n(x)= 0$.\\ 
\hline   
\end{tabular}
\end{table}
\end{itemize}

\subsubsection{Abstraction and Modeling}

Abstraction is a way of expressing an idea in a specific context while at the same time suppressing details irrelevant in that context. In other words, abstraction is a process that extracting the common features and ignoring details from concrete entities. Abstraction connects a concrete problem and its generalized version, which share some common patterns. The conceiving, designing, implementation and operation of the solution to the very general problem plays an active and essential role in problem-solving. 

Modeling is a reasonable approximation of the reality, which focuses on the main factors and eliding minor issues. Usually, abstraction and modeling is closely related. When modeling, reasonable approximation should be considered carefully, which controls the measure of errors and the level of precision of the solution. For computing the CEI-1, there are different methods and parameters for specifying the errors of approximation. 

For determining the period of MP, there are several abstraction and modeling processes with different levels. We summarize them as follows: 
\begin{itemize}
\item Physical abstraction and mathematical modeling. \Tab \ref{tab-abstract-model-phys-ma} shows the physical abstraction forms and corresponding mathematical modeling encountered in this paper. In this table, physical abstraction emphasizes the nature of physics while the mathematical model emphasizes the quantitative description, which reflects the relation of physics and mathematics. 
      \begin{itemize}
      \item The semi-quantitative method based on the $\Pi$ theorem of dimension analysis is a special approach which catches the essential relation of physics quantities and ignoring some constants.
      \item The quantitative method based on the linear approximation in the sense of harmonic oscillator leads to a simple mathematical model which can be understood by college students and even students in senior middle school with the help of Newton's second law of motion and Hooke's law about equivalent system of spring-mass. 
      \item The quantitative method based on the conservation law of energy is instructive for students with physics intuition and sufficient knowledge of calculus.   
      \item In exploring the formulae of period for MP, the fundamental abstraction-modeling are the particle model and rigid body model with an ideal rotation, which include the following assumptions: ignoring the shape and size of the small ball in the pendulum system; ignoring the variation of the radius of the pendulum system (thus the trajectory is an arc of a circle); ignoring the air resistance; ignoring the motion out of the vertical plane, or equivalently ignoring the Coriolis effect.

\begin{table*}[htb]
\centering
\caption{Physical Abstraction and Mathematical Modeling ($x$ is the angular displacement)}
\label{tab-abstract-model-phys-ma}
\resizebox{\textwidth}{!}{
\begin{tabular}{lll}
\hline
\textbf{Type}       & \textbf{Physical Abstraction}  & \textbf{Mathematical Modeling} \\
\hline \hline
Semi-quantitative ($\Pi$ theorem)   &   Dimension analysis           		    & $ [T] = [m]^\alpha [g]^\beta [\ell]^\gamma$
 $\Longrightarrow \physdim{T} = \physdim{M}^\alpha \physdim{L}^\beta\physdim{T}^{-2\beta} \physdim{L}^\gamma$
\\
\hline
Quantitative (linear model)        &   Harmonic oscillator & $\displaystyle E = m\ell^2\dot{x}^2 + \frac{1}{2}mg\ell x^2$  
$\displaystyle \Longrightarrow \ddot{x}+\frac{g}{\ell}x = 0$ \\
\hline
Quantitative  (nonlinear model)        & Conservation of energy  &  
$E = m\ell^2\dot{x}^2 + mg\ell (1-\cos x)$ \\
\qquad Particle & \qquad Motion of particle & 
$\displaystyle{v = \ell\dot{x}, f = m\dot{v}}$ 
$\Longrightarrow m\ell \ddot{x} = -mg\sin x $ \\ 
\qquad Rigid body & \qquad Rotation of rigid body & 
$I = m\ell^2, \omega = \dot{x}, M = I\dot{\omega}$ 
$\Longrightarrow m\ell^2 \ddot{x} = -mg\ell \sin x$\\
\hline
\end{tabular}
}
\end{table*}
\end{itemize}
\item For the simple linear model of the MP, the fundamental assumption is the small amplitude $\theta$ such that $\sin \theta \approx \theta$. In this case, the approximation 
$K(k)\approx K(0) = \pi/2$
is taken. An alternative approximation is also possible by taking $(1+x)^\alpha\approx 1 + \alpha x$ for $\alpha \in \Real$ and small $x$, which implies that $\frac{1}{\sqrt{1-k^2\sin^2u}}= (1-k^2\sin^2u)^{-1/2}
\approx 1 + \frac{1}{2}k^2\sin^2u$. Therefore,
\begin{equation}
\begin{split}
K(k) &= \int^{\pi/2}_0 \frac{\dif u}{\sqrt{1-k^2\sin^2u}}\\
     &\approx \int^{\pi/2}_0 \left[1+ \frac{1}{2}k^2\sin^2u \right]\dif u \\
     &=\frac{\pi}{2}\left(1+ \frac{1}{4}k^2 \right)
\end{split}
\end{equation}
Obviously, this is the second order approximation of $K(k)$ when it is compared with \eqref{eq-K-series}.
\item For the CEI-1 obtained, the abstraction-modeling involved in four different solvers are summarized in \Tab \ref{tab-abs-mod}. 

	\begin{table*}[htb]
	\centering
	\caption{Mathematical Abstraction and Modeling for Computing CEI-1}
	\label{tab-abs-mod}
	\resizebox{\textwidth}{!}{
	\begin{tabular}{lll}
	\hline
	\textbf{Solver} & \textbf{Abstraction} (Problem generalization) & \textbf{Modeling} (Reasonable approximation) \\
      \hline  \hline     
      IS-CEI-1  & $\displaystyle{K(k) \to S(x) = \sum^\infty_{m=0}c_mx^m}$  & $S(x)\approx S_m(x), \abs{S_{m+1}(x)-S_m(x)}< \varepsilon$ \\
      \hline
      AGM-CEI-1 & $\displaystyle{K(k) \to \AGM(a,b)}\to \vec{x} =\scrd{\mathcal{A}}{agm}(\vec{x}) $  &  $\vec{x}\approx \vec{x}_{i+1} =\scrd{\mathcal{A}}{agm}(\vec{x}_i), \quad \norm{\vec{x}_{i+1}-\vec{x}_i}<\varepsilon $ \\
      \hline
      GC-CEI-1  & $\displaystyle{K(k) \to \int^a_{-a}\frac{f(x,k)}{\sqrt{a^2 -t^2}}\dif x }$  &   $\displaystyle{\int^a_{-a}\frac{f(x,k)}{\sqrt{a^2-x^2}}\dif x =\int^1_{-1} \frac{f(at, k)}{\sqrt{1-t^2}}\dif t \approx \sum_{i=0}^{n-1} \scru{w}{gc}_i f(a\scru{x}{gc}_i, k)}$ \\
      & $\displaystyle{\sum_{i=0}^{n-1}w_if(ax_i,k)\to \braket{\vec{w}}{\vec{v}_f(k,a)}}$ &   $\abs{\scru{\mathcal{E}}{gc}_n(f,a,k)}<\varepsilon$\\
      \hline
      GL-CEI-1  & $\displaystyle{K(k)  \to \int^b_a f(x,k)\dif x }$  &  $\displaystyle{\int^b_{a}f(x,k)\dif x =\frac{b-a}{2}\int^1_{-1}f\left(u(t),k\right)\dif t\approx \frac{b-a}{2}\sum_{i=0}^{n-1} \scru{w}{gl}_i f(\scru{u}{gl}_i,k) }$  
      \\
      & $\displaystyle{\sum_{i=0}^{n-1}w_if(x_i,k,a,b)\to \braket{\vec{w}}{\vec{v}_f(k,a,b)}}$ &       $ u(t) = \cfrac{b+a+t(b-a)}{2}, \quad \abs{\scru{\mathcal{E}}{gl}_n(f,a,b,k)}<\varepsilon$\\
              &  $P_n(x) = 0 \to f(x,\alpha) = 0$    & $\displaystyle{x_{i+1} = \scrd{\mathcal{A}}{Newton}(x_i,f,\alpha),\quad  \abs{x_{i+1} - x_i}<\varepsilon}$ \\
       & $(P_{n+1}, P_n, P_{n-1}) \to (Y_{n+1}, Y_n, Y_{n-1}) $   &  \\
       & $(P'_{n+1}, P'_n, P'_{n-1}) \to (Y'_{n+1}, Y'_n, Y'_{n-1}) $   &  \\
      \hline
      \end{tabular}
}
      \end{table*}            
\end{itemize}

It should be remarked that abstraction helps us to solve more general problems which have more applications and generic solutions instead of the specific problems encountered. 

\subsubsection{Logical and Algorithmic Thinking}

Both logic and algorithms are essential to STEM, CDIO and CT. They appear in step of solving practical problems, particularly in the C-D-I steps in the interaction of CDIO and CT, see \Fig \ref{fig-interaction-CDIO-CT}. Although humans have innate and intuitive understanding of logic and algorithms, the two core concepts are mathematical in nature. Both logic and algorithm are precise and systematic, which means that intuition is not enough to capture their features and many mistakes may happen in a wide variety of situations or under various conditions. 

There are two fundamental types of logic reasoning: inductive and deductive reasoning. In order to instruct computers to make logical decisions, the Boolean logic are widely used in designing algorithms and computer programs.    
For the numerical computation problem of solving CEI-1 with algorithms, the stopping condition is based on Boolean logic and the rules of controlling computational errors. Mathematically, the Cauchy's criterion, i.e., $\abs{S_{m+1}(x)-S_m(x)} < \varepsilon$, $\norm{\vec{x}_{i+1} - \vec{x}_i} < \varepsilon$ or $\abs{x_{i+1}-x_i}<\varepsilon$ is used for computing the infinite series and the fixed-points of interest.

Algorithmic thinking follows abstraction and modeling directly. Once the concrete problem has been generalized and the modeling process has been determined, the algorithmic thinking will be activated for making a sequence of clearly defined steps to solve the generic problem. Roughly speaking, each subproblem decomposed will correspond to an algorithm, which is capable of solving a class of problems and can be reused many times.

\subsubsection{Anticipating and Dealing with Bugs}


The word \textit{bug} is used to represent a fault in a solution that can cause erroneous behavior in computer science and engineering. 
Bugs can be captured and eliminated at any stage in the development or research process, but the
earlier you find and fix them the better profits you get. Without doubt, it is much easier to change something 
when it exists in the conceiving and/or designing process rather than after the implementation process. Getting rid of bugs is a good practice in solving all kinds of problem, especially complex problems.

For computing the CEI-1, there exist some typical potential bugs behind the whole project.  \Tab \ref{tab-cope-bugs} lists the potential bugs and corresponding strategies in solving the CEI-1.
\begin{table*}[htp]
\centering
\caption{Anticipating and Coping with Bugs/Errors}
\label{tab-cope-bugs}
\resizebox{\textwidth}{!}{
\begin{tabular}{lll}
\hline
\textbf{Subproblem}  &\textbf{Possible Bug}  & \textbf{Strategy} \\
\hline \hline
$\displaystyle{\frac{(2m-1)!!}{(2m)!!}}$   &  $n!!$ overflows   &    Replacing $\displaystyle{\frac{(2m-1)!!}{(2m)!!}}$ 
by $\displaystyle{ \prod^m_{i=1}\left[1-\frac{0.5}{i} \right]}$\\
\hline
$ \displaystyle{\frac{\abs{S_{m}(x)-S_{m-1}(x)}}{\abs{S_m(x)}}} \overset{?}{<} \varepsilon $   & $S_m(x) = 0$  &  
$\displaystyle{\frac{\abs{S_{m}(x)-S_{m-1}(x)}}{\abs{S_m(x)}+\delta}}\overset{?}{<} \varepsilon$  for $\delta = 10^{-10}$\\
\hline
$\displaystyle{x_{i+1} = x_i -\frac{f(x_i,\alpha)}{f'(x_i,\alpha)}}$
   & $f'(x_i,\alpha) = 0$  &  Checking the slope $f'(x_i, \alpha)$ when computing $x_{i+1}$ with $x_i, f(x_i, \alpha)$ and $f'(x_i,\alpha)$ \\
\hline
\end{tabular}
}
\end{table*}

\subsubsection{Verification, Validation and Testing}

For the solution obtained to the given problem, there is a necessary stage that verifies, validates and tests the very solution. This stage is known as verification-validation-testing (VVT) or simply verification and validation (V \& V). Evaluating the solution is a fundamental step in the sense of computational thinking as well as system engineering.

For the sub-problems decomposed in solving the period of MP and algorithms developed in this paper, we can construct some simple but useful test cases to verify the solution to the computational tasks involved.

As an application of VVT, the testing and evaluating programs is widely applied in software developing or computer programming.
It is a standard engineering method that is introduced in computational thinking. Usually, programs consists of modules. 
The modules designed should be tested carefully in order to verify their functions and performances. We can test the functions of the procedures by binding appropriate parameters and comparing the output with the results for verification in the table.

It should be noted that commercial software can be used in VVT process to test and evaluate our own programs. However, we should keep an eye on the commercial software and do not depend on them blindly. Actually, for the verification of our algorithms for computing CIE-1 $K(k)$, we find that the numerical solutions to the
$K(k)$ for $k\in \set{0.00, 0.10, 0.20, 0.30, 0.40, 0.50}$ by \lstinline|EllipticK| in  Mathematica \cite{WolframEllipIntegral} and \lstinline|ellipticK| in MATLAB are \textit{not} acceptable \cite{Zhang2022-CEI1}. 

\begin{table}[htp]
\centering
\caption{A comparison of numerical solvers for $K(k)$}
\label{tab-comparison-K-solvers}
\resizebox{0.8\textwidth}{!}{
\begin{tabular}{cccc}
\hline
\makecell[c]{\textbf{Parameter} \\ $\quad$ \\ $k$ }  & \makecell[c]{\textbf{CEI-1 Solvers} \\ $\ProcName{Kseries/Kagm}$ \\ $\ProcName{Kintgc/Kintgl}$ } & \makecell[c]{\textbf{MATLAB}\\ function \\  \lstinline|ellipticK| } & \makecell[c]{\textbf{Mathematica} \\ function \\ \lstinline|EllipticK|} \\
\hline
$0.00$  &    1.5707963    &    1.5708     &  1.57080  \\
$0.10$  &    1.5747456    &    1.6124     &  1.61244  \\
$0.20$  &    1.5868678    &    1.6596     &  1.65962  \\
$0.30$  &    1.6080486    &    1.7139     &  1.71389  \\
$0.40$  &    1.6399999    &    1.7775     &  1.77752  \\
$0.50$  &    1.6857504    &    1.8541     &  1.85407  \\ 
\hline
\end{tabular}
}
\end{table}   

\begin{figure*}[htbp]
\centering
\subfigure[Output of \lstinline|EllipticK| in Mathematica]{
\boxed{\includegraphics[height=2cm]{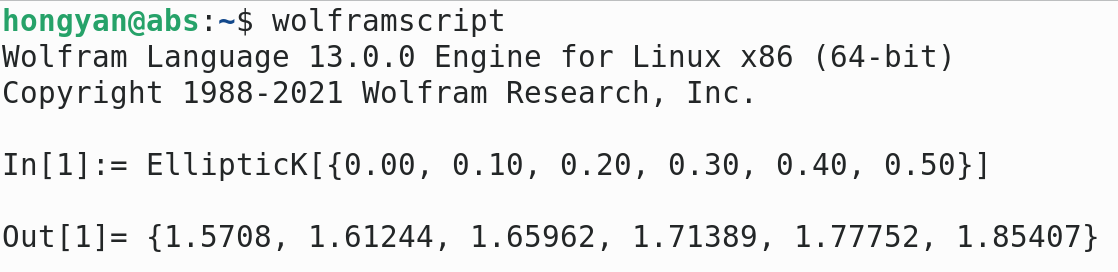}} 
}
\subfigure[Output of \lstinline|ellipticK| in MATLAB]{
\boxed{\includegraphics[height=2cm]{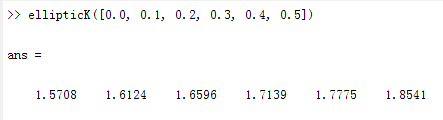}} 
}
\caption{Unacceptable solutions via commercial software Mathematica and MATLAB}
\end{figure*}

It should be emphasized that it is unwise to repudiate the commercial software. What we should do is to show the significance of VVT for the students and guide them to follow the right way to exploring the solution to the problem. In such a reliable way, we can help the students to avoid possible losses in the future. 

\subsection{Problem-Project-Oriented Learning for Solving the Period of MP}

The problem of solving the period of MP can be used to train college or graduate students with the PDT method via team cooperation. It is feasible to organize a team or group with $3\sim 4$ students to work on a project for solving the period of MP with one specific method such as infinite series method, AGM method, Gauss-Chebyshev method or Gauss-Legendre method. It is also possible for the team to solve the problem with all of the four methods discussed above and explore the advantages and disadvantages behind them.

\subsubsection{Framework for Solving the Period of MP}

The framework shown in \Fig \ref{fig-team-coop} is useful for the instructor, he/she can figure out the key issues for the students clearly. For the teacher, his/her attention could be put on the following perspectives:
\begin{itemize}
\item Explaining the objective clearly. Firstly, it is necessary to explain the background of MP and the task of finding the period given an arbitrary initial angle amplitude. Secondly, it is important to explain the CDIO based on how to do and  CT via how to think. Thirdly, it is the key issue that a combination of CDIO and CT is essential for solving complex STEM problem. 
\item Converting the problem step by step with multi-disciplines according to \Fig \ref{fig-MP-STEM}. The conversion demonstrates the four elements ( viz. S, T, E, and M) in STEM and decomposition of problem in the sense of CT.
\item Helping students in conceiving the solution to the period of MP. It is necessary to instruct the students to exploring the formulae of period from simple to complex. The three fundamental modeling methods --- dimension analysis, linear approximation and general nonlinear model --- should be interpreted  briefly and clearly.
\item Explaining the definition of CEI-1  and the four numerical schemes shown in \Fig \ref{fig-ways2K} for computing the $K(k)$ with sufficient details. 
\item Giving supplementary interpretation of generating orthogonal polynomials $Y_n(x)$  and their derivatives $Y'_n(x)$ such that 
\begin{equation} \label{eq-ortho}
\left\{
\begin{split}
Y_{n}(x) &= (A_n x + B_n)Y_{n-1}(x) - C_n Y_{n-2}(x), n\ge 2  \\
Y_0(x)   &= c_0\\
Y_1(x)   &= c_1 x + c_2
\end{split}
\right.
\end{equation}
via iterative method and solving the roots of Legendre polynomials $P_n(x)$ with Newton-Raphson method.
\item Interpret and emphasize the interaction of CDIO and CT in the process of PPOL.
\end{itemize}

\subsubsection{Potential Difficulties for Students}

For the purpose of exploring the solution to solving the period of MP, there are two potential difficulties for students. 

\begin{figure}[htb]
\centering
\includegraphics[width=0.8\textwidth]{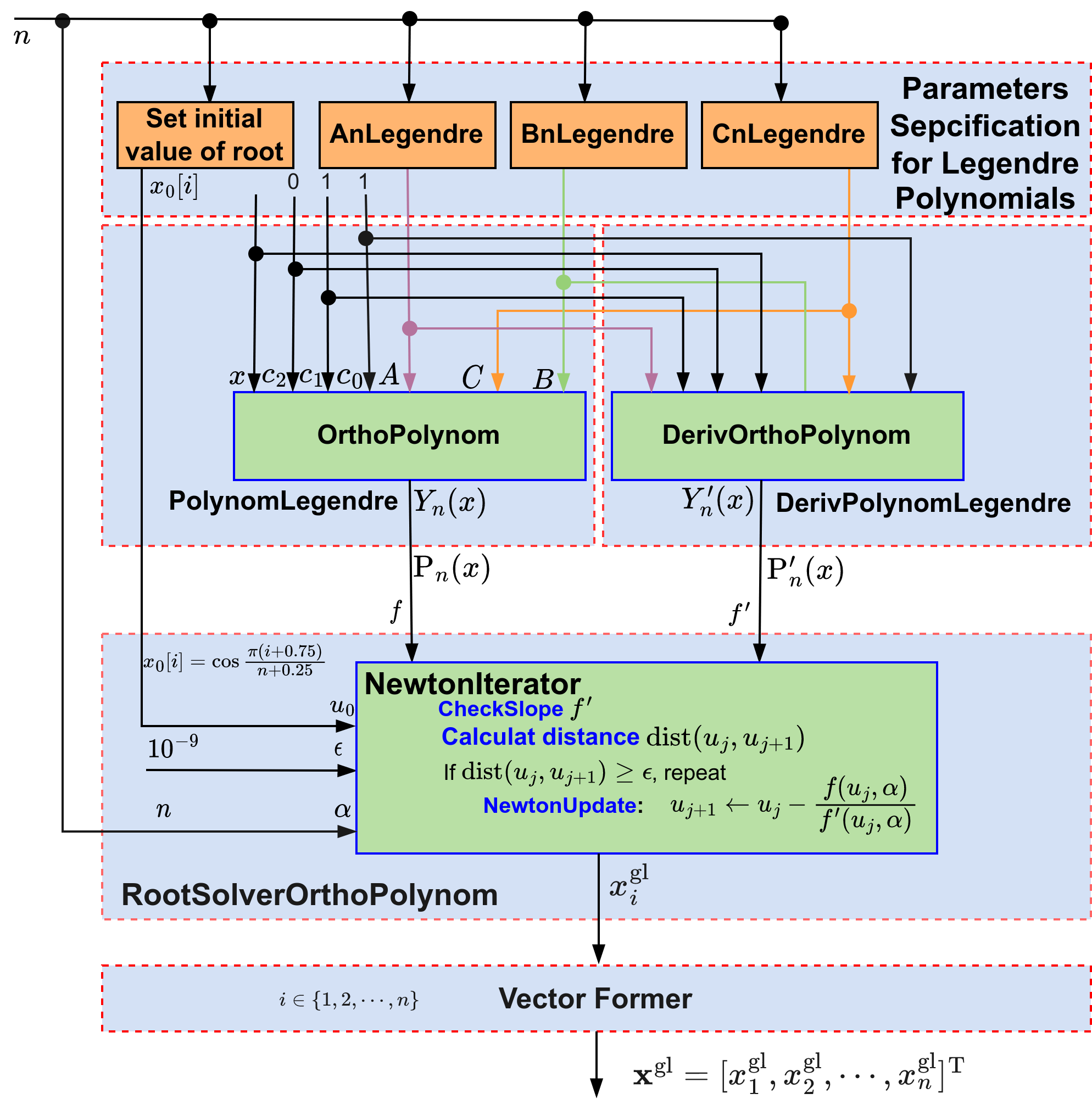} 
\caption{Solving the root vector $\scru{\vec{x}}{gl}$ for the Legendre polynomial $P_n(x)$ such that  $P_n(\scru{x}{gl}_{i})=0$.}
\label{fig-NewtonSolver}
\end{figure}

The first difficulty may arise in finding the root-weight pairs for $P_n(x)$   for arbitrary $n\in \mathbb{N}$. \Fig \ref{fig-NewtonSolver} demonstrates the system of solving the root vector of $P_n(x)$.  Usually in the books about special mathematical functions \cite{CourantHilbert-vol1,SJZhang1996,HandbookMPES2011}, the different kinds of orthogonal polynomials, such as the Legendre polynomials $P_n(x)$, Chebyshev polynomials $T_n(x)$ of the first or second kind, Laguerre polynomials $L_n(x)$ and Hermite polynomials $H_n(x)$,  are treated separately, which hides the general iterative formulae \eqref{eq-ortho}. Once the $i$-th root $\scru{x}{gl}_i$ is obtained, the corresponding weight $\scru{w}{gl}_i$ can be determined by \eqref{eq-weight-Legendre}.

\begin{figure*}[htb]
\centering
\includegraphics[width=\textwidth]{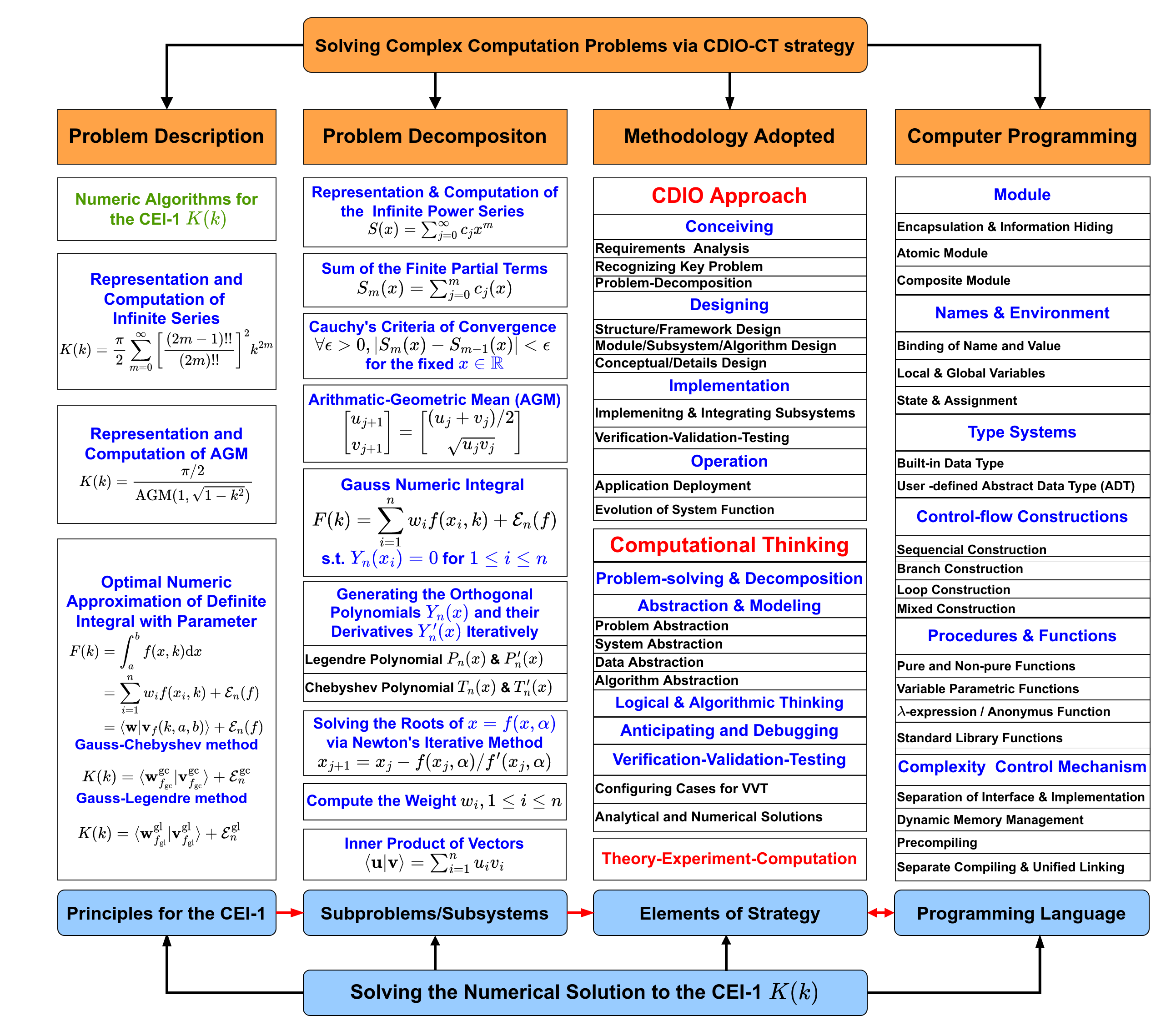} 
\caption{A global view of computing the CEI-1 with the  CDIO-CT collaborative strategy}
\label{fig-comp-CEI-1}
\end{figure*}

The second difficulty may lie in establishing the global view of solving the numerical solution to the CEI-1 with the CDIO-CT strategy. Although there are four optional methods, there are gaps  between the problem description and computer programming when designing algorithms and coding. \Fig \ref{fig-comp-CEI-1} illustrates the important factors for solving complex computation problems via CDIO-CT strategy: problem description, problem decomposition, methodology adopted and computer programming.  For solving the numerical solution to the CEI-1 $K(k)$, the problem decompositions of the four numerical schemes are listed clearly, which can be used for constructing subsystems in designing numerical algorithms. For more details, please see Zhang et al \cite{Zhang2022-CEI1}.

\subsection{Data Collection and Analysis}

It is necessary to collect and analyze the data about the students' work on the project of solving complex STEM problem in order to check the training effect and obtain feedback for further improvement.

\subsubsection{Data Collection}

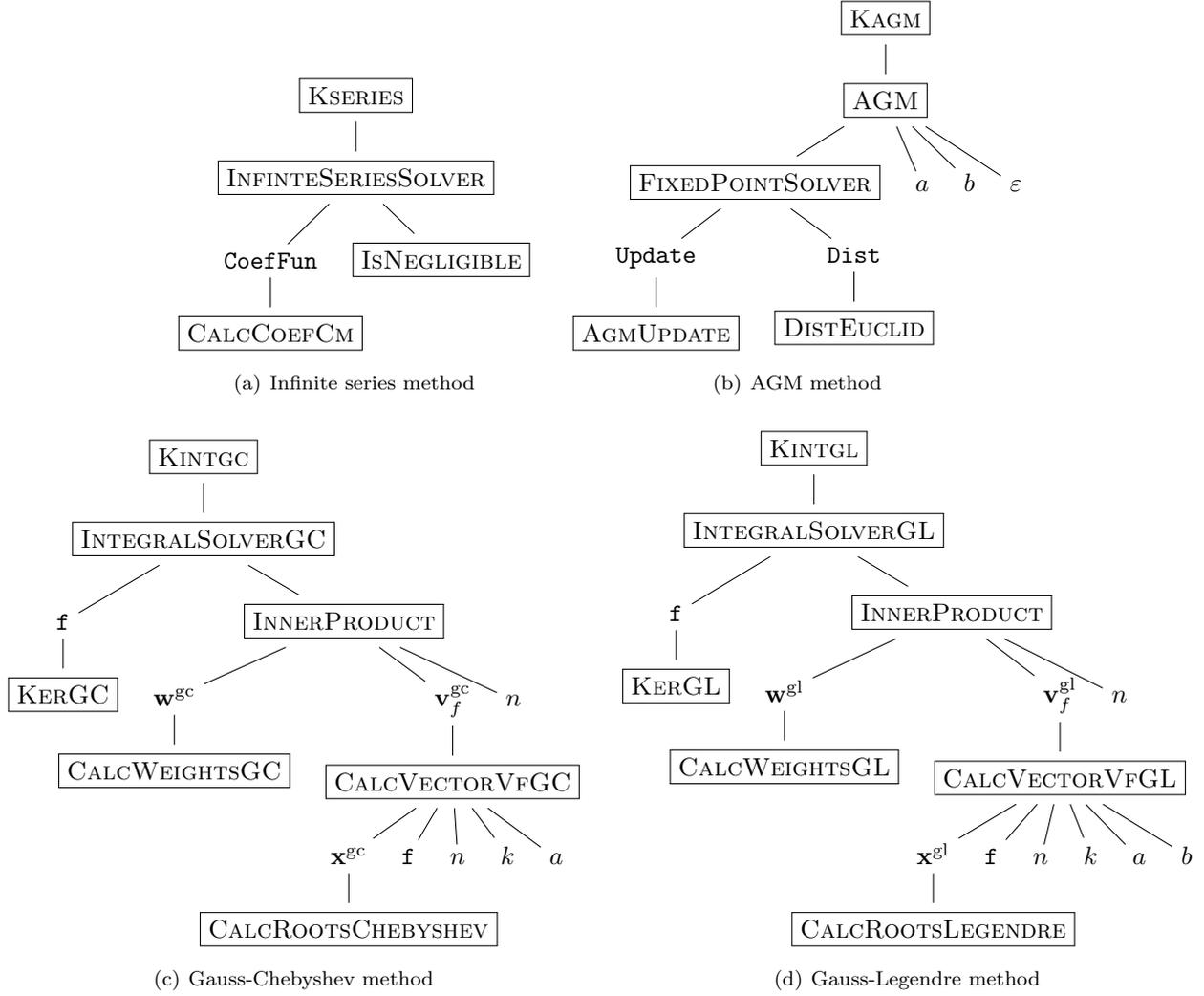
\begin{figure*}[htb]
\centering
	\subfigure[Infinite series method]{
	\begin{forest}
		[\fbox{\ProcName{Kseries}}
		[\fbox{\ProcName{InfinteSeriesSolver}}
		[\PtrToFun{CoefFun}[\fbox{\ProcName{CalcCoefCm}}]]
		[\fbox{\ProcName{IsNegligible}}]
		]
		] 
	\end{forest}
	}
	\subfigure[AGM method]{
		\begin{forest}
		[\fbox{\ProcName{Kagm}}
		[\fbox{\ProcName{AGM}}
		[\fbox{\ProcName{FixedPointSolver}}
		[\PtrToFun{Update}[\fbox{\ProcName{AgmUpdate}}]]
		[\PtrToFun{Dist}[\fbox{\ProcName{DistEuclid}}]]
		]
		[$a$]
		[$b$]
		[$\varepsilon$]
		]
		] 
	\end{forest}
	} \\
	\subfigure[Gauss-Chebyshev method]{
		\begin{forest}
		[\fbox{\ProcName{Kintgc}}
		[\fbox{\ProcName{IntegralSolverGC}}
		[\PtrToFun{f}[\fbox{\ProcName{KerGC}}]]
		[\fbox{\ProcName{InnerProduct}}
		[$\scru{\vec{w}}{gc}$
		[\fbox{\ProcName{CalcWeightsGC}}]    	
		]
		[$\scru{\vec{v}}{gc}_f$
		[\fbox{\ProcName{CalcVectorVfGC}}
		[$\scru{\vec{x}}{gc}$
		[\fbox{\ProcName{CalcRootsChebyshev}}]
		]
		[\PtrToFun{f}]	
		[$n$]
		[$k$]	
		[$a$]
		]
		]
		[$n$]
		] 
		]	
		] 
	\end{forest}
	}
	\subfigure[Gauss-Legendre method]{
		\begin{forest}
		[\fbox{\ProcName{Kintgl}}
		[\fbox{\ProcName{IntegralSolverGL}}
		[\PtrToFun{f}[\fbox{\ProcName{KerGL}}]]
		[\fbox{\ProcName{InnerProduct}}
		[$\scru{\vec{w}}{gl}$[\fbox{\ProcName{CalcWeightsGL}}]]
		[$\scru{\vec{v}}{gl}_f$
		[\fbox{\ProcName{CalcVectorVfGL}}
		[$\scru{\vec{x}}{gl}$
		[\fbox{\ProcName{CalcRootsLegendre}}]	        
		]
		[\PtrToFun{f}]	
		[$n$]
		[$k$]
		[$a$]
		[$b$]
		]	
		]
		[$n$]
		] 
		]	
		] 
	\end{forest}
	}
	\caption{Modules for computing CEI-1: \ProcName{Kseries}, \ProcName{Kagm}, \ProcName{Kintgc} and \ProcName{Kintgl}}
	\label{fig-design-interface}
\end{figure*}

The data collection stage includes the following terms involved in the cycle of the project:
\begin{itemize}
\item Data in conceiving. It is necessary for the students to outline the possible ways of solving the STEM problem of interest according to the literature and the hints from the instructor. For solving the period of MP, the semi-quantitative method, the simple quantitative method and complex quantitative method should be discussed and the expression of period $T$ should be recorded. The steps for deriving the formula $T=\lambda \sqrt{\ell/g}$ should be recorded by the students with draft block diagram. 
The \Tab \ref{tab-T-lambda} and \Fig \ref{fig-MP-STEM} can be used for checking the team work for the instructor. 
\item Data in the description of essential problem. The philosophy of STEM education implies there are various equivalent descriptions for the problem. Each description corresponds to a feasible solution. The team should state the 
 description clearly, record it with formulae, diagrams or sentences and submit the description to the instructor. \Fig \ref{fig-ways2K} is helpful for the instructor to check the students' description of CEI-1.
\item Data in problem decomposition. The specified essential problem can be decomposed into sub-problems according to its structure of the problem with the computational thinking. Students should record the structure of the essential problem and state the sub-problems clearly. The second column of \Fig \ref{fig-comp-CEI-1}, \Tab \ref{tab-PD-IS}, \Tab \ref{tab-PD-AGM}, \Tab \ref{tab-PD-GC-CEI} and \Tab \ref{tab-PD-GL-CEI} are helpful for the instructor to check the decomposition schemes in different teams.
\item Data in designing. The modules for solving the sub-problems can be designed separately once the framework and interfaces are defined clearly. 
Students should submit the outlines of the modules, interfaces as well as the solving methods to the instructor. 
For computing the CEI-1, \Fig \ref{fig-design-interface} is helpful for checking the interfaces of the modules for solving the essential problem with four equivalent forms and sub-problems. For more details about the algorithms, please see Zhang et al \cite{Zhang2022-CEI1}. 
\item Data in implementation. The data for implementation includes the details about hardware, software and so on. For computing the CEI-1, the program code for the algorithms is essential. It should be noted that the algorithms can be implemented with various programming languages. The version controlling can be 
done with the Git conveniently. 
\item Data in summary for the project. The project report could be written with \LaTeX{} or Word by the team members. The project report is important for the evaluation of the project.  
\end{itemize}

\subsubsection{Data Analysis}

The data collected in the management of project are usually not structured and there is no unified way for data analysis. It is necessary for the instructor to extract abstract information so as to  compare the learning effects and provide feedback for students. A fundamental rule for validating the data collected is the step of VVT in the implementation.

\section{Results} 
\label{sec-result}

\subsection{Global View of Computing CEI-1 with CDIO-CT Collaborative Strategy}

In this paper, we take the problem of solving the period of MP and conceiving the solution with STEM analysis, system modeling and simulation. Our analysis shows that solving the numerical solution to the CEI-1, viz. $K(k)$,  is the key issue of exploring the period of MP in the sense of general nonlinear model. Computing the CEI-1 is a typical complex STEM problem, which involves science (physics), technology (computer science and software design), engineering (software engineering and system engineering), and mathematics (calculus, numeric analysis,  linear algebra, and functional analysis) respectively.

\begin{figure}[htb]
	\centering
	\includegraphics[width=0.8\textwidth]{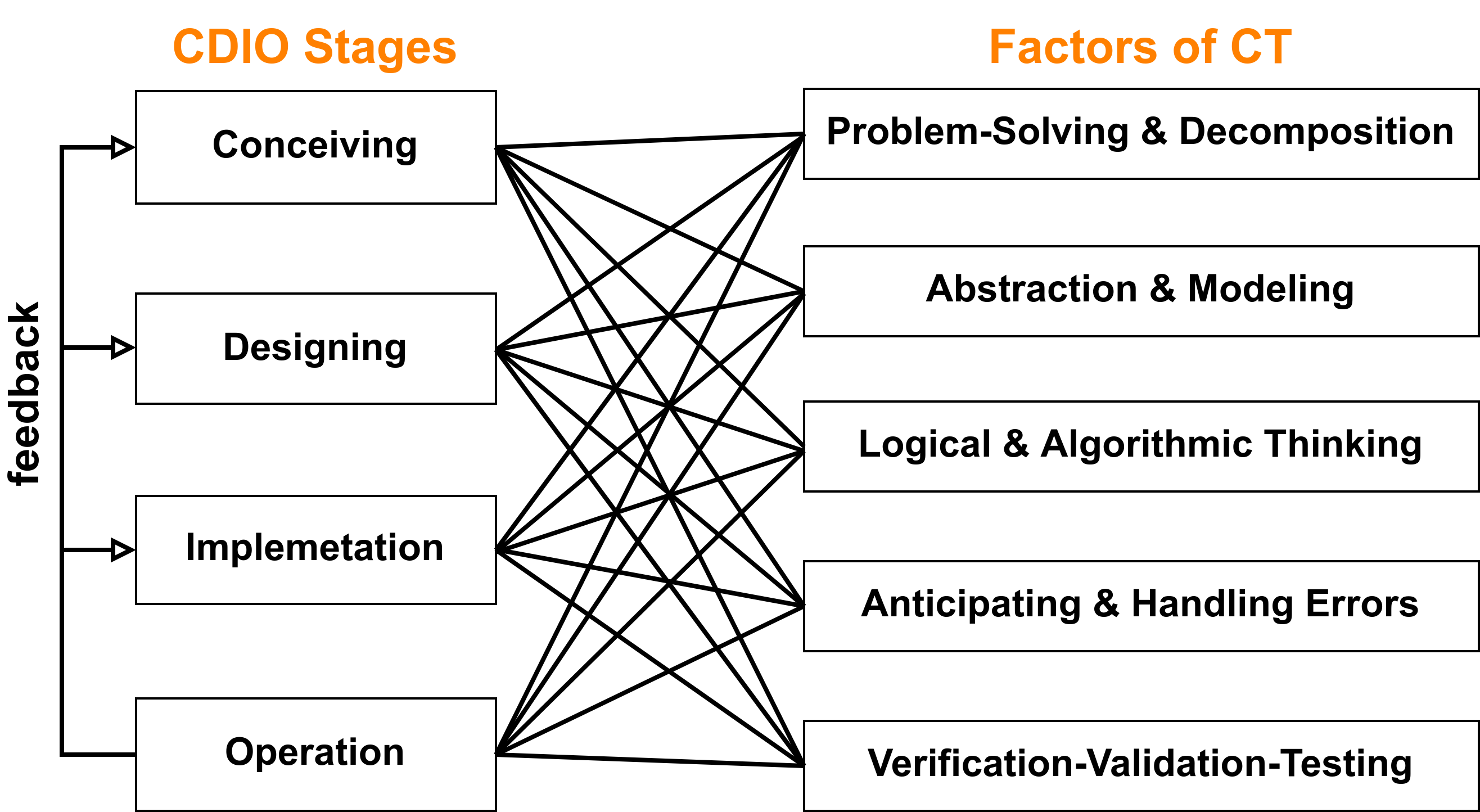} 
	\caption{Interaction of CDIO stages and CT}
	\label{fig-interaction-CDIO-CT}
\end{figure}

 Our methodology is a combination of CDIO approach and CT, which interprets the research paradigm of triplet of theory-experiment-computation through a well-designed interdisciplinary project of solving the period of MP. \Fig \ref{fig-interaction-CDIO-CT} illustrates the interaction of CDIO approach and CT. It is obvious that the four stages of CDIO process are interwoven with the five factors of CT. 

The methodology adopted and tools required for exploring the period of MP are connected with ach other. \Fig \ref{fig-comp-CEI-1} shows a global view of computing the CEI-1. In the level of STEM education, solving complex computation problem via CDIO-CT strategy includes four parts: problem description, problem decomposition, methodology adoption and computer programming. In the level of tools for solving the numerical solution to the CEI-1 $K(k)$, the problem description outlines the principle for computing $K(k)$, the problem decomposition leads to sub-problems or subsystems, the methodology implies elements of strategy, and the computer programming is implemented via programming language. The routine of training students' ability of solving complex STEM problems can be summarized as follows:
\begin{itemize}
	\item describing the problem concerned clearly;
	\item decomposing the problem into sub-problems properly;
	\item adopting the CDIO-CT collaborative strategy to explore the solutions to the sub-problems and merge them into the objective solution; and
	\item exploring the relations of solutions and tools required.
\end{itemize}  

For solving the period $T = \lambda \sqrt{\ell/g}$ of MP, there are three fundamental ways in the sense of physics science:
\begin{itemize}
	\item the semi-quantitative method via dimension analysis shows that $\lambda$ is a unknown constant to be measured with an experiment;
	\item the linear approximation method with the assumption of small oscillation implies that $\scrd{\lambda}{linear} = 4K(0) = 2\pi$;
	\item the nonlinear method states that $\scrd{\lambda}{nonlinear} = 4K(k)$. 
\end{itemize} 
There are four feasible methods for computing the CEI-1 $K(k)$:
\begin{itemize}
	\item infinite series method,
	\item AGM method,
	\item Gauss-Chebyshev method, and
	\item Gauss-Legendre method.
\end{itemize}
The sub-problems decomposed are summarized as follows by recognizing patterns and using abstraction:
\begin{itemize}
	\item representation and computation of the infinite power series,
	\item sum of the finite partial terms,
	\item Cauchy's criteria of convergence,
	\item the iterative process in AGM,
	\item the Gauss numeric integral,
	\item the generation of orthogonal polynomials $Y_n(x)$ and their derivatives $Y'_n(x)$ by iterative method and specifying specific initial values,
	\item solving the roots of nonlinear equation with Newton's method,  
	\item the computation of the weights appearing in the Gauss Numeric integration, and finally
	\item the calculation of inner product of two $n$-dim vectors.    
\end{itemize}
These sub-problems are direct generalization of the specific sub-tasks of computing the value $K(k)$. The solutions to these sub-problems have wide applications in various disciplines depending on the practitioners' professional backgrounds.    

Our CDIO-CT collaborative strategy demonstrates the four stages of CDIO approach, five factors of CT and their interactions clearly: 
\begin{itemize} 
	\item On the one hand, the conceiving stage relies on the perspective of STEM, which reduces the solving of period into  computing the CEI-1 and four equivalent solutions to the computational task. In the designing stage, we construct a series of algorithms for the general sub-problems and primitive problem. For the implementation stage, we presents 
	the C code by separating the interfaces and implementation details. Meanwhile, all of the algorithms developed are verified by creating optional test cases. The operation is briefly mentioned by combining theory and experiment. 
	\item On the other hand, the factors of CT are discussed one by one with the achievements in the CDIO process. Almost all of the general rules of CT can be extracted from the process of CDIO. We should emphasize that the abstract and modeling are vital to building a system, which includes \textit{problem abstraction}, \textit{system abstraction}, \textit{data abstraction} and \textit{algorithm abstraction}. As stimulated by the unacceptable results of computing the CEI-1 by MATLAB and Mathematica, we can teach students a fundamental fact: \textit{we should keep an eye on commercial software and the process of VVT is necessary}, otherwise we may obtain a troublesome solution.  
\end{itemize} 

The solution to the period of MP is obtained via computer programming, which depends on programming language and techniques closely.        
There are some common features to be considered carefully for the implementation of computing CEI-1 with some programming language such as C and Python. By working on exploring the period of MP, the students will benefit from the following prospects:
\begin{itemize}
\item deepening the understanding of CDIO approach, CT and their interaction;
\item building a vivid example of solving complex STEM problem by combining physics, mathematics, computer science, software engineering and system engineering;
\item obtaining profound experiences about computer programming,  such as modules, names \& environments, type systems, control-flow constructions, procedure \& functions, and complexity control mechanism;
\item improving team collaboration capabilities which includes project division, joint writing, discussion, using tools such as \LaTeX{} and Git, skills of presentation,  and so on.  
\end{itemize} 

\subsection{General Framework for Solving Complex STEM Problem in Modeling and Simulation}

\begin{figure*}[htb]
\centering
\includegraphics[width=\textwidth]{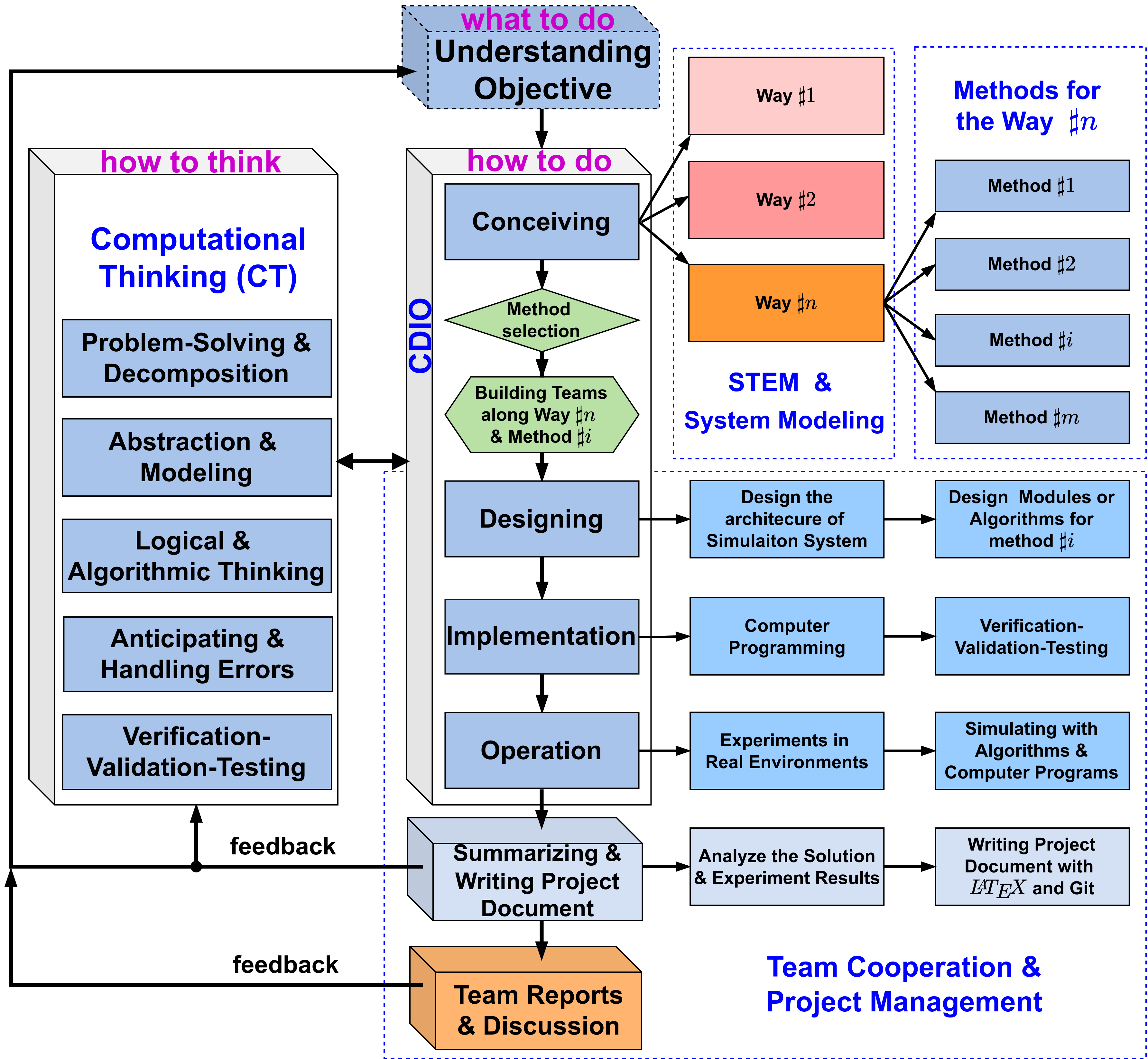} 
\caption{General framework for solving complex STEM problems in modeling and simulation}
\label{fig-general-framework}
\end{figure*}

Although the  framework shown in \Fig \ref{fig-team-coop} is designed for 
the project of solving the period of MP, it can be extended to general complex STEM problems and projects in system modeling and simulation. In \Fig \ref{fig-general-framework}, there is more than one way for the STEM problem and multiple methods for the ways concerned. The ways proposed in conceiving are 
based on different principles with different complexity and levels of effectiveness. For the way with best approximation of modeling the problem of interest and best effectiveness of solving the problem, its essence is usually the most complex. By comparison, we can find that: the dimension analysis, linear approximation and nonlinear modeling in solving the period of MP have been replaced by the ways with different complexity; the infinite series method, the AGM method, Gauss-Chebyshev method and Gauss-Legendre method in the specific framework have been generalized to the different methods in the general framework.   

\section{Discussion} \label{sec-discussion}

\subsection{Beyond the Computation of CEI-1}

Although there are three kinds of complete elliptic integral, only the first kind of complete elliptic integral $K(k)$ is discussed in this paper. The ideas of computing $K(k)$ with four methods can also be applied to computing of the second and third kinds of complete elliptic integral \cite{SJZhang1996}. 

The dimension analysis is widely used in fluid mechanics and aerodynamics. Its generalization is the type checking in computer science and software engineering. In dimension analysis, the essence of the $\Pi$ theorem for balancing the dimension can be described by linear algebra, which leads to simple linear algebraic system with unique solution.  

The system and modules shown in \Fig \ref{fig-NewtonSolver} for solving the roots of Legendre polynomial $P_n(x)$ can be modified and reused for various applications. With different setting of $c_0, c_1, c_2, A_n, B_n, C_n$ in \eqref{eq-ortho} and initial value $x_0[i]$ for the $i$-th root, we can obtain  other orthogonal polynomials and their  derivatives as well as roots.

The code released on GitHub for computing CEI-1 $K(k)$ is written in the C programming language. The algorithms proposed in this paper can be implemented with other high level computer programming languages such as Python, MATLAB, Octave, C++, Java, Julia and so on.  For the beginners of learning computer programming, they will be more sensitive on the grammars of computer programming language than the general principles. In consequence, it will be better if some teacher assistants are available for the students. 

For example, it should be noted that the significant digits should be considered when working on the project of exploring the period of MP. In the sense of system modeling and simulation, the value of $K(k)$ can be computed with high precision, say $\epsilon = 10^{-9}$, and the effects of the precision of $g$ and $\ell$ can be ignored. However, if the project for solving the period of MP is taken as a STEM project and practical measurement of $g$ and $\ell$ are performed, then we have to deal with the significant digits for the values of $\scrd{g}{meas}$, $\scrd{\ell}{meas}$ and $\scrd{T}{meas}$ where the subscript ``meas'' means the result of measurement. In this sense, the period will be 
\begin{equation}
\scrd{T}{simu} = 4K(k)\sqrt{\frac{\scrd{g}{meas}}{\scrd{\ell}{meas}}}
\end{equation}  
which implies that it is pointless if we calculate $K(k)$ with a high precision. In other words, if there are 
2 effective digits for $g$ and $\ell$, then we should set the precision $\epsilon = 10^{-2}$ when computing $K(k)$ with infinite series method and AGM method, and the parameter $n\in \mathbb{N}$ in Gauss-Chebyshev method and Gauss-Legendre method could be set with a smaller number. In consequence, only two significant digits in \Tab \ref{tab-comparison-K-solvers} are worth  considering in the sense of science.

As an improvement of MP, C. Huygens proposed the concept of symmetric isochronous pendulum.  Recently, J. M. Randazzo et al.  proposed the asymmetric  isochronous pendulum for the first time \cite{Randazzo2018,Cidoncha2020}.  The discussion of isochronous pendulum is an extension of this work on MP and the computation of the period will be a new evolution, which could be a new project for the students. 

\subsection{Limitation}

As illustrated in the metaphor shown in \Fig \ref{fig-road-map}, the idea of the CDIO-CT collaborative strategy for problem-project-oriented STEM education is universal. However, the general framework   shown in \Fig \ref{fig-general-framework} is limited for solving the complex STEM problem in system modeling and simulation. For those out of the scope, the framework should be modified according to the constrains encountered. For the STEM problems arising in K12, the general framework can be simplified according to the practical problems encountered. 

\section{Conclusions} \label{sec-conclusions}

A general framework for solving complex STEM problems with the CDIO-CT collaborative strategy is proposed by demonstrating the interdisciplinary project of solving the period of MP. In the sense of global perspective, the CDIO approach is about \textit{``how to do''}, the CT is about \textit{``how to think''} and the problem to be solved is about \textit{``what to do}.  Our analysis shows that the CDIO stages and the elements of CT are interwoven closely.
Although the teaching method for the STEM education is with three different names, viz. problem-based learning, project-based learning and project-driven teaching,  it is integrated in the general framework naturally.

The philosophy of STEM education is that each problem has more than one way to the feasible solutions. The project about solving the period of MP provides a good example for practicing this philosophy. 
The concepts and tools arising in developing the software for calculating CEI-1 are valuable for teaching computer programming. The methodology embedded in the project of exploring the expression and solution to the period of MP is enlightening, which is worth recommending to students and instructors in colleges and universities.

\section*{Data Availability Statement}

The C code for the implementation can be downloaded from the following  GitHub site 

\begin{center}

\textcolor{blue}{\href{https://github.com/GrAbsRD/CEI-1/tree/main}{https://github.com/GrAbsRD/CEI-1/tree/main}}

\end{center}

\section*{Acknowledgment}

The first author's thanks go to Prof. Qi Zhang and Prof. Yan Han for the training and discussion of the CDIO approach in Civil Aviation University of China. The authors' thanks go to Prof. Li-Hua Wu and Dr. Mi Wang for their suggestions for the manuscript of this paper. The authors' thanks also go to Juan M. Randazzo for his comment on the history and concepts of symmetric/asymmetric isochronous pendulum.

This work was supported by the
Hainan Provincial Education and Teaching Reform Project of Colleges and Universities under grant number Hnjg2019-46, and in part by the 
Hainan Provincial Natural Science Foundation of China 
under grant number 720RC616, and in part by the National Natural Science Foundation of China under grant number
62167003.

\noindent\textbf{How to cite this article}: \\
H.‐Y. Zhang, Y. Zhou,Y.‐T. Li, F.‐Y. Li, and Y.‐H. Jiang, CDIO‐CT
collaborative strategy for solving complex STEM
problems in system modeling and simulation: an
illustration of solving the period of mathematical
pendulum, \textit{Computer Applications in Engineering  Education}, (2023),
e22698. \url{https://doi.org/10.1002/cae.22698}

\end{document}